\documentclass{osa-article}

\journal{osajournal} 
\usepackage{caption}
\usepackage{subcaption}
\usepackage{float}
\usepackage{soul} %for strikethrough feature
\usepackage{graphicx}
\newcommand{\squeezeup}{\vspace{-2mm}}
\usepackage[normalem]{ulem}
\usepackage{amsmath}
\usepackage{tabularx}
\newcommand{\stkout}[1]{\ifmmode\text{\sout{\ensuremath{#1}}}\else\sout{#1}\fi}
\articletype{Research Article}
\begin{document}
\title{Sub-MHz spectral dip in a resonator-free twisted gain medium}
\author{Neel Choksi\authormark{1*}, Yi Liu\authormark{1}, Rojina Ghasemi\authormark{1}, Li Qian \authormark{1*}} %and Author Three\authormark{2,3}}
\address{\authormark{1}Department of Electrical and Computer Engineering, University of Toronto, 10 Kings College Road, Toronto, Ontario M5S 3G4, Canada}
%\\
%\authormark{2}Publications Department, The Optical Society (OSA), 2010 Massachusetts Avenue NW, Washington, DC 20036, USA\\
%\authormark{3}Currently with the Department of Electronic Journals, The Optical Society (OSA), 2010 Massachusetts Avenue NW, Washington, DC 20036, USA}
%\email{\authormark{*}xyz}
\email{\authormark{*}neel.choksi@mail.utoronto.ca, l.qian@utoronto.ca} %% email address is required



\begin{abstract*}
Ultra-narrow optical spectral features resulting from highly dispersive light-matter interactions are essential for a broad range of applications such as spectroscopy, slow-light, and high-precision sensing. Features approaching sub-MHz, or equivalently, Q-factors approaching ~1 billion and beyond, are challenging to obtain in solid-state systems, ultimately limited by loss. We present a novel approach to achieve tunable sub-MHz spectral features, at room temperature, without resonators. We exploit gain-enhanced polarization pulling in a twisted birefringent medium where polarization eigenmodes are frequency-dependent. Using Brillouin gain in a commercial spun fiber, we experimentally achieve a 0.72 MHz spectral dip, the narrowest backward Brillouin scattering feature ever reported. Further optimization can potentially reduce the linewidth to $<$0.1 MHz. Our approach is simple and broadly applicable, offering on-demand tunability and high sensitivity, with a wide range of applications such as microwave photonic filters, slow and fast light, and optical sensing.
\end{abstract*}

%%%%%%%%%%%%%%%%%%%%%%%%%%  body  %%%%%%%%%%%%%%%%%%%%%%%%%%
\section*{Introduction}
Ultra-narrow resonances are highly sought-after for a wide range of applications such as slow-light \cite{tsakmakidis2017ultraslow,vigneron2020loss} and information storage \cite{kim2015non,dong2015brillouin,wang2020electromagnetically}, high precision sensing \cite{arora2018high}, microwave photonics \cite{choudhary2017high,wen2018ultrahigh,liu2020integrated,marpaung2019integrated}, spectroscopy \cite{suh2016microresonator}, frequency stabilization \cite{anderson2018highly}, light detection and ranging (LIDAR) \cite{suh2018soliton,trocha2018ultrafast}, and optical gyroscopes \cite{li2017microresonator, liang2017resonant,lai2020earth}. Narrowing the linewidth further to sub-MHz range (or equivalently, increasing the Q factors to approaching a billion and beyond) is very challenging, but highly desirable, as it greatly improves the key performance metrics, across several applications \cite{vahala2003optical}. For example, when a sub-MHz transparency is induced via Brillouin scattering using a high-Q silica resonator \cite{kim2015non}, it produces slow-light with a 5-orders-of-magnitude higher delay bandwidth product than the previous record reported for Brillouin-based systems. Such favourable performance improvements have stimulated phenomenal progress in obtaining narrow resonances through platforms such as gas-phase atomic systems \cite{sayrin2015storage, gouraud2015demonstration}, photonic crystal cavities \cite{savchenkov2004kilohertz,grudinin2006ultrahigh,anderson2018highly,liu2018low}, whispering gallery mode (WGM) \cite{shitikov2018billion,wu2020greater}, and microring resonators \cite{ji2017ultra, yang2018bridging,puckett2021422}, slow-light fiber Bragg gratings (FBGs) \cite{vigneron2020loss}, and phase-shifted FBGs \cite{jing2017impedance}. However, despite significant progress, there remain many challenges with these platforms. 

While narrow spectral features can be realized in gas-phase atomic systems \cite{sayrin2015storage, gouraud2015demonstration}, they typically require low temperatures and low pressure vacuums, making them bulky, expensive and impractical. Solid-state systems such as photonic crystal cavities \cite{savchenkov2004kilohertz,grudinin2006ultrahigh,anderson2018highly,liu2018low}, WGM \cite{shitikov2018billion,wu2020greater} and microring resonators \cite{ji2017ultra, yang2018bridging,puckett2021422} are much more preferable, but to realize sub-MHz features, complex and costly fabrication steps like chemical-mechanical polishing and high temperature annealing are required \cite{chang2020ultra}, which are not sufficiently mature to allow large-scale commercial production of ultrahigh Q devices. Specialty FBGs, such as $\pi$-phase-shifted FBGs have also been shown to exhibit narrow resonances \cite{jing2017impedance}, but their linewidths are limited to a few MHz due to their intrinsic losses. Recently, efforts have been made to narrow the linewidth of a slow-light FBG \cite{vigneron2020loss} further, by using a gain medium to offset the intrinsic loss. But to achieve the narrowest resonance of 8.5 fm ($\sim$1.1 MHz), the FBGs need to be probed with very low signal powers ($<$–50 dBm), limiting its practical use due to low signal-to-noise ratios. 

Stimulated Brillouin scattering (SBS) is another technique to realize narrow spectral features. Both forward and backward SBS are possible, though forward SBS involves transverse acoustic modes, and is far weaker than backward SBS in conventional single-mode waveguides \cite{kobyakov2010stimulated}. While sub-MHz features have been demonstrated using forward SBS in an ultrahigh-Q micro-resonator \cite{kim2015non,dong2015brillouin}, achieving and maintaining triple resonance (one acoustic and two optical resonances) in an ultrahigh Q resonator is no easy feat. Backward SBS, on the other hand, is easily obtainable in conventional waveguides, but its linewidth is a few tens of MHz in most solids \cite{eggleton2019brillouin, kobyakov2010stimulated}. Recent efforts made to narrow the SBS feature further include: making a microwave analogue of electromagnetically induced absorption (EIA) \cite{yelikar2020analogue} using destructive interference, and combining SBS with an ultrahigh Q resonator \cite{wen2018ultrahigh}. However, as with other SBS-resonator combination experiments, \cite{wen2018ultrahigh} requires precise alignment between the SBS peak and one of the resonances of the ultrahigh-Q resonator. Obtaining and maintaining such alignment over time can be a major challenge. Moreover, the configuration becomes significantly more complex with the added resonator.

In this work, we propose an entirely new paradigm to realize a sub-MHz, tunable spectral feature without using any resonator. We exploit gain-enhanced polarization pulling in a twisted birefringent medium with frequency-dependent polarization eigenmodes. To demonstrate a specific realization, we use backward SBS in a commercial spun birefringent fiber (SBF) and experimentally achieve a 0.72 MHz spectral dip, which is to our knowledge, the narrowest backward SBS spectral feature ever reported. We also demonstrate the on-demand tunability of the linewidth, depth, and the spectral location of this dip. Furthermore, the configuration is extremely simple, no different from a conventional SBS arrangement, and no need for resonator alignment or stabilization, as it is resonator-free. 

\section*{Theoretical Framework}
An SBF is an elliptically birefringent fiber, fabricated by spinning a birefringent preform while drawing the fiber. It is characterized by its twist rate ($k_t=\frac{2\pi}{L_t}$) and its unspun linear birefringence ($k_{b} = \frac{2\pi}{L_b}$)\cite{przhiyalkovsky2017polarization}, where $L_t$ and $L_b$ are the twist period and the beat length of the SBF, respectively. 
Note, while $L_t$ and $k_t$ are independent of the frequency ($\nu$) of the optical field, $L_b$ is \textit{dependent} on $\nu$, and so is $k_b$. To denote $L_b$ and $k_b$ for signal and pump fields, we use the notation $L_b(\nu_{s/p})$ and $k_b(\nu_{s/p})$ throughout the paper. Here s and p denote the signal and pump fields, respectively.

\begin{figure}[H]
    \centering
    \includegraphics[scale=0.6]{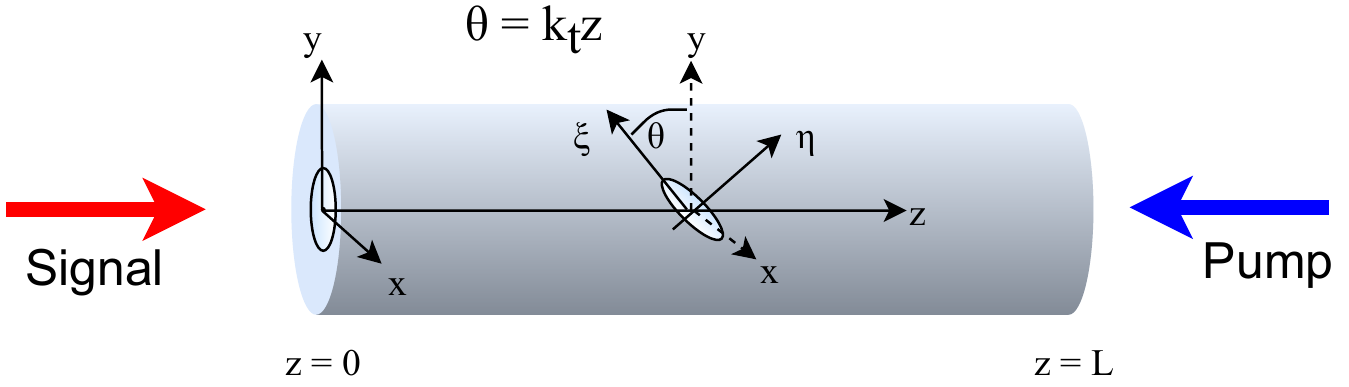}
    \caption{The fixed coordinate system ($x$,$y$,$z$) and the right-handed rotating coordinate system ($\xi,\eta,z$) are used for modelling SBS in spun fiber. The ($\xi,\eta,z$) coordinates are aligned with the fast and slow axes of the fiber locally, and rotating with $z$ at the twist rate $k_t$.}
    \label{fig:sbf rotating frame}
\end{figure}

An SBF has two \textit{{elliptically}} polarized eigenmodes \cite{przhiyalkovsky2017polarization,yeh2009optics,feldman1993polarization}, which, unlike conventional birefringent fibers (i.e., polarization-maintaining fibers or PMF), are dependent on both the twist (spun) period and the beat length of the SBF, and are therefore frequency-dependent \cite{zhu2010phase,feldman1993polarization,yeh2009optics}. Based on an SBS model for single-mode fiber (SMF) \cite{stolen1979polarization,boyd2019nonlinear,agrawal2000nonlinear,zadok2008vector}, we develop an SBS model for SBF, for the configuration shown in Fig. \ref{fig:sbf rotating frame}. The signal field is launched into SBF at $z = 0$, while the pump is launched at $z = L$, where $L$ is the length of SBF. The twist is modelled using a right-handed rotating frame of reference whose coordinates ($\xi,\eta,z$) are aligned with the fast and slow axes of the fiber locally. Hence, the ($\xi,\eta,z$) coordinates are rotating with $z$ at the twist rate $k_t$. The $(x,y,z)$ coordinate system is considered to be the fixed frame of reference, and without loss of generality, we assume that at $z = 0$, the fixed frame of reference is aligned with the rotating frame of reference.

In the ($\xi,\eta,z$) coordinate system, the non-normalized Jones vector for the signal field propagating in $+z$ direction is denoted by $\Vec{A}_{rs}(z)$, and for the pump field, counter-propagating in $-z$ direction, by $\Vec{A}_{rp}(z)$. The amplitude of $\Vec{A}_{r}(z)$ corresponds to the square root of the optical power in the mode. The input signal power and input pump power are denoted by $P_{sig}$ and $P_{pump}$, respectively. The signal frequency ($\nu_s$) is downshifted from the pump frequency ($\nu_p$) by a frequency $\nu$ ($\nu_s$ = $\nu_p-\nu$). The evolution of pump and signal field amplitudes $\Vec{A}_{rs}(z)$ and $\Vec{A}_{rp}(z)$ along the fiber is described by a set of coupled complex vector equations:
\begin{align}
    \frac{d\Vec{A}_{rs}(z)}{dz} &=  \frac{d\boldsymbol{T}_s(z)}{dz}\boldsymbol{T}^\dagger_s(z)\Vec{A}_{rs}(z) + \gamma(\Omega,\Omega_B)(\Vec{A}_{rp})\Big((\Vec{A}_{rp})^\dagger(z).\Vec{A}_{rs}(z)\Big) - \frac{\alpha}{2}\Vec{A}_{rs}(z)
        \label{Stokes with coupled mode}\\
    \frac{d\Vec{A}_{rp}(z)}{dz} &=  \frac{d\boldsymbol{T}_p(z)}{dz}\boldsymbol{T}^\dagger_p(z)\Vec{A}_{rp}(z) + \gamma(\Omega,\Omega_B)(\Vec{A}_{rs})\Big((\Vec{A}_{rs})^\dagger(z).\Vec{A}_{rp}(z)\Big) + \frac{\alpha}{2}\Vec{A}_{rp}(z) \label{Pump with coupled mode}
\end{align}
In Eqs. \eqref{Stokes with coupled mode} and \eqref{Pump with coupled mode}, the first term represents the polarization rotation in SBF without gain, the second term represents the SBS term, and the third term is the fiber loss. Here, $\gamma(\Omega,\Omega_B)$ [W m$]^{-1}$ is the Brillouin gain coefficient, which is modelled by a Lorentzian function, given by:
\begin{equation}
    \gamma(\Omega,\Omega_B) = \gamma_0 \frac{(\Gamma_B/2)^2}{(\Omega_B - \Omega)^2+(\Gamma_B/2)^2},
\end{equation}
where $\gamma_0$ is the line-center gain factor, $\Gamma_B$ is the bandwidth of the SBS gain, $\Omega_B$ ($\Omega_B = 2\pi\nu_B$) is the Brillouin frequency shift, and $\Omega$ ($\Omega=2\pi\nu$) is the frequency of the acoustic wave. For the simulations presented in this paper, the value of $\gamma_0$ is chosen to be 0.896 [W m$]^{-1}$ (on the same order as the values stated in \cite{boyd2019nonlinear,agrawal2000nonlinear,nikles1997brillouin}). $\boldsymbol{T}_{s/p}$ is the polarization transfer matrix for the signal (subscript $s$) and the pump (subscript $p$) field in the ($\xi,\eta,z$) coordinate system, given by \cite{mcintyre1978light,sakai1981birefringence,huang1998microwave}:
\begin{equation}
    \boldsymbol{T}_{s/p}(z) = \begin{bmatrix}
    cos(k_{bts/p}z) - j\frac{k_{bs/p}}{k_{bts/p}}sin(k_{bts/p}z)& -\frac{k_t}{k_{bts/p}}sin(k_{bts/p}z)\\
    \frac{k_t}{k_{bts/p}}sin(k_{bts/p}z)& cos(k_{bts/p}z) + j\frac{k_{bs/p}}{k_{bts/p}}sin(k_{bts/p}z)
    \end{bmatrix}, \label{transform matrix secf} 
\end{equation} 
where $k_{bts/p} = \sqrt{k_{bs/p}^2 + k_t^2}$. The eigenmodes of the transfer matrix $\boldsymbol{T}_{s/p}(z)$ \cite{yeh2009optics,feldman1993polarization,przhiyalkovsky2017polarization} are denoted by $\boldsymbol{S_1}$ ($\boldsymbol{P_1}$) and $\boldsymbol{S_2}$ ($\boldsymbol{P_2}$) at the signal (pump) frequency, and they are given by:

\begin{subequations}
  \begin{tabularx}{\textwidth}{Xp{0.005cm}X}
  \begin{equation}
     \boldsymbol{S_1}= \frac{1}{m_s}\begin{bmatrix}i\Big(\frac{k_{bs}+k_{bts}}{k_t}\Big)\\1\end{bmatrix}\label{S_1}
  \end{equation}
  & &
  \begin{equation}
    \boldsymbol{S_2}= \frac{1}{m'_s}\begin{bmatrix}i\Big(\frac{k_{bs}-k_{bts}}{k_t}\Big)\\1\end{bmatrix}\label{S_2}
  \end{equation}\\
 \begin{equation}
     \boldsymbol{P_1}= \frac{1}{m_p}\begin{bmatrix}i\Big(\frac{k_{bp}+k_{btp}}{k_t}\Big)\\1\end{bmatrix}\label{P_1}
  \end{equation}
 & &
  \begin{equation}
    \boldsymbol{P_2}= \frac{1}{m'_p}\begin{bmatrix}i\Big(\frac{k_{bp}-k_{btp}}{k_t}\Big)\\1\end{bmatrix}\label{P_2}
  \end{equation}
  \end{tabularx}
\end{subequations}
where $m_{s/p}$ and $m'_{s/p}$are the normalizing factors. These eigenmodes are independent of the fiber position ($z$), meaning that polarization eigenmodes launched into the fiber are maintained throughout the fiber. Conversely, non-eigenmode input will experience polarization change along the fiber, even under passive conditions (see the dotted and solid green curve in Fig. \ref{fig:Polarization pulling}c). As can be verified, the eigenmodes $\boldsymbol{S_1}$ and $\boldsymbol{P_2}$, are nearly orthogonal, and so are $\boldsymbol{S_2}$ and $\boldsymbol{P_1}$ (Refer to the blue and red dots on the Poincar\'e sphere in Figs. \ref{fig:Polarization pulling}a and \ref{fig:Polarization pulling}b). Here we use "nearly" because the pump and signal frequencies are slightly different, and therefore the eigenmode for the pump is not exactly orthogonal to the eigenmode for the signal. To quantify the amount of overlap between signal and pump polarization, we define a polarization overlap factor ($F$)
\begin{equation}
    F(z) = \Bigg|\frac{\Vec{A}^\dagger_{rs}(z).\Vec{A}_{rp}(z)}{|A_{rs}(z)||A_{rp}(z)|}\Bigg|^2. \label{overlap as a function of length}
\end{equation}
The polarization overlap factor ($F$) remains very small ($<10^{-11}$, see the solid and dashed red curve in the inset of Fig. \ref{fig:Polarization pulling}c) when we launch orthogonal eigenmodes $\boldsymbol{S_1}(\nu_s)$ and $\boldsymbol{P_2}(\nu_p)$.

When signal power at frequency $\nu_s$ is launched into the eigenmode $\boldsymbol{S_1}(\nu_s)$ without the pump (i.e., no SBS gain), its polarization is maintained along the fiber (see the dashed red curve in the inset of Fig. \ref{fig:Polarization pulling}c). But, when pump power is launched into $\boldsymbol{P_2}(\nu_p)$ eigenmode, the signal field experiences a weak polarization pulling effect (see the red trace in Fig. \ref{fig:Polarization pulling}a and the red curve in Fig. \ref{fig:Polarization pulling}c) because of its small overlap $F$ with the pump polarization.

The polarization eigenmodes of a twisted birefringent medium (e.g., SBF) are frequency-dependent. $\boldsymbol{S_1}(\nu_s)$ is an eigenmode for $\nu_s$, but it is not an eigenmode for $\nu'_s$. Note that, when a non-eigenmode is launched into SBF, its polarization is not preserved along the length of SBF, even under passive conditions (see the dotted and solid green curve in Fig. \ref{fig:Polarization pulling}c). So, with a minute deviation in signal frequency from $\nu_s$ to $\nu'_s$, its polarization is no longer an eigenmode for that frequency ($\nu'_s$), resulting in a periodic polarization variation along the fiber. This can be seen from the oscillation in $F$ (dotted and solid green curve in Fig. \ref{fig:Polarization pulling}c). This oscillation causes the signal to be more susceptible to polarization pulling \cite{zadok2008vector,millot2014nonlinear,stiller2012demonstration} induced by the pump. Consequently, the signal experiences more SBS gain, which in turn causes its polarization to deviate further, leading to more polarization pulling. Such positive feedback, analogous to laser action, ultimately pulls the signal polarization (at $\nu'_s$) significantly towards the pump (Fig. \ref{fig:Polarization pulling}b), resulting in a significant increase in SBS gain. This explains why the SBS gain increases drastically away from the dip frequency (at $\nu_s$).
\begin{figure}[H]
\centering
\begin{subfigure}{0.7\textwidth}
   \centering
   \includegraphics[width=\textwidth]{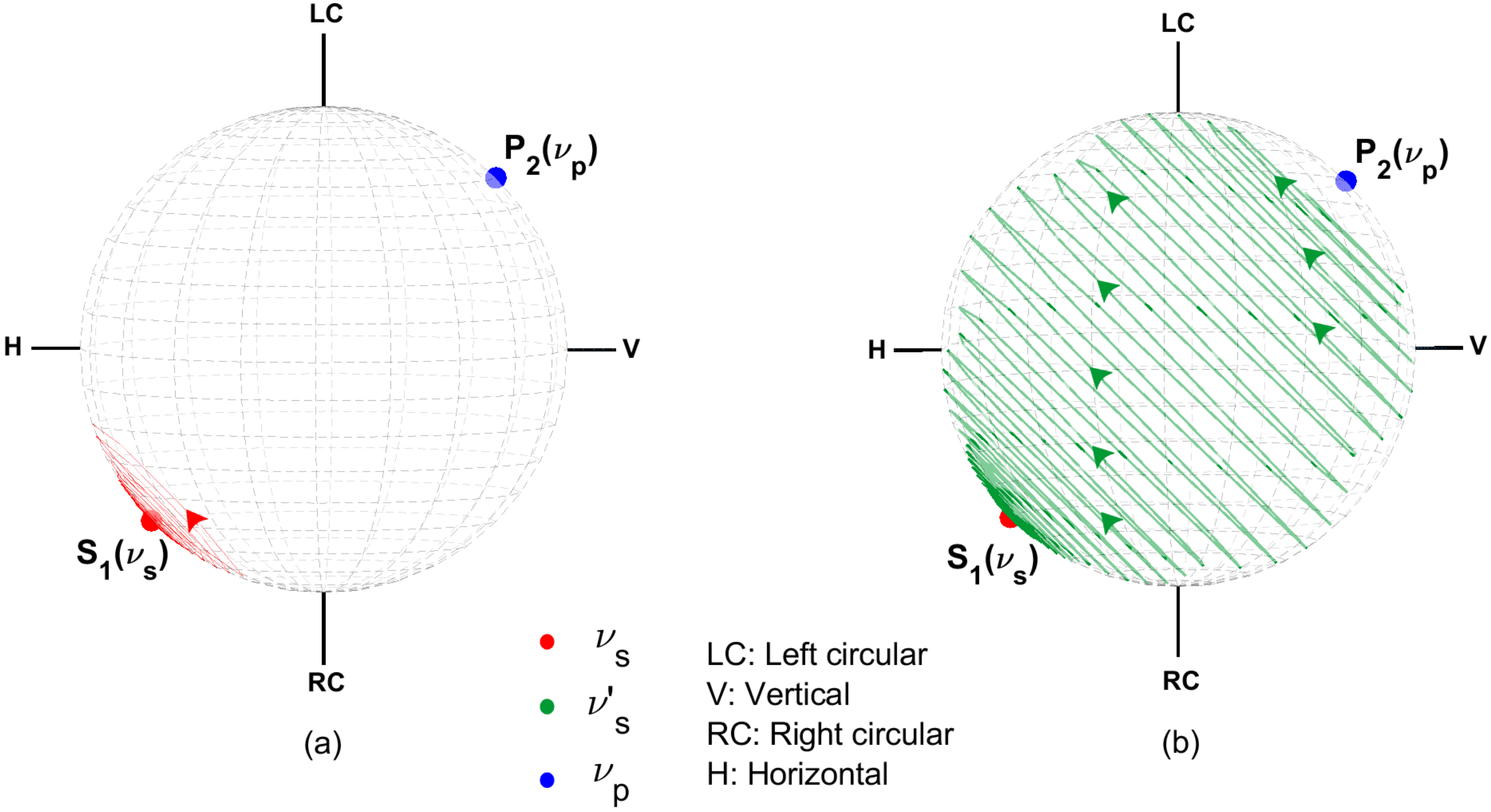}
   \caption*{}
   \label{fig:poincare polarization pulling}
\end{subfigure}
\begin{subfigure}{0.7\textwidth}
\centering
\includegraphics[width=\textwidth]{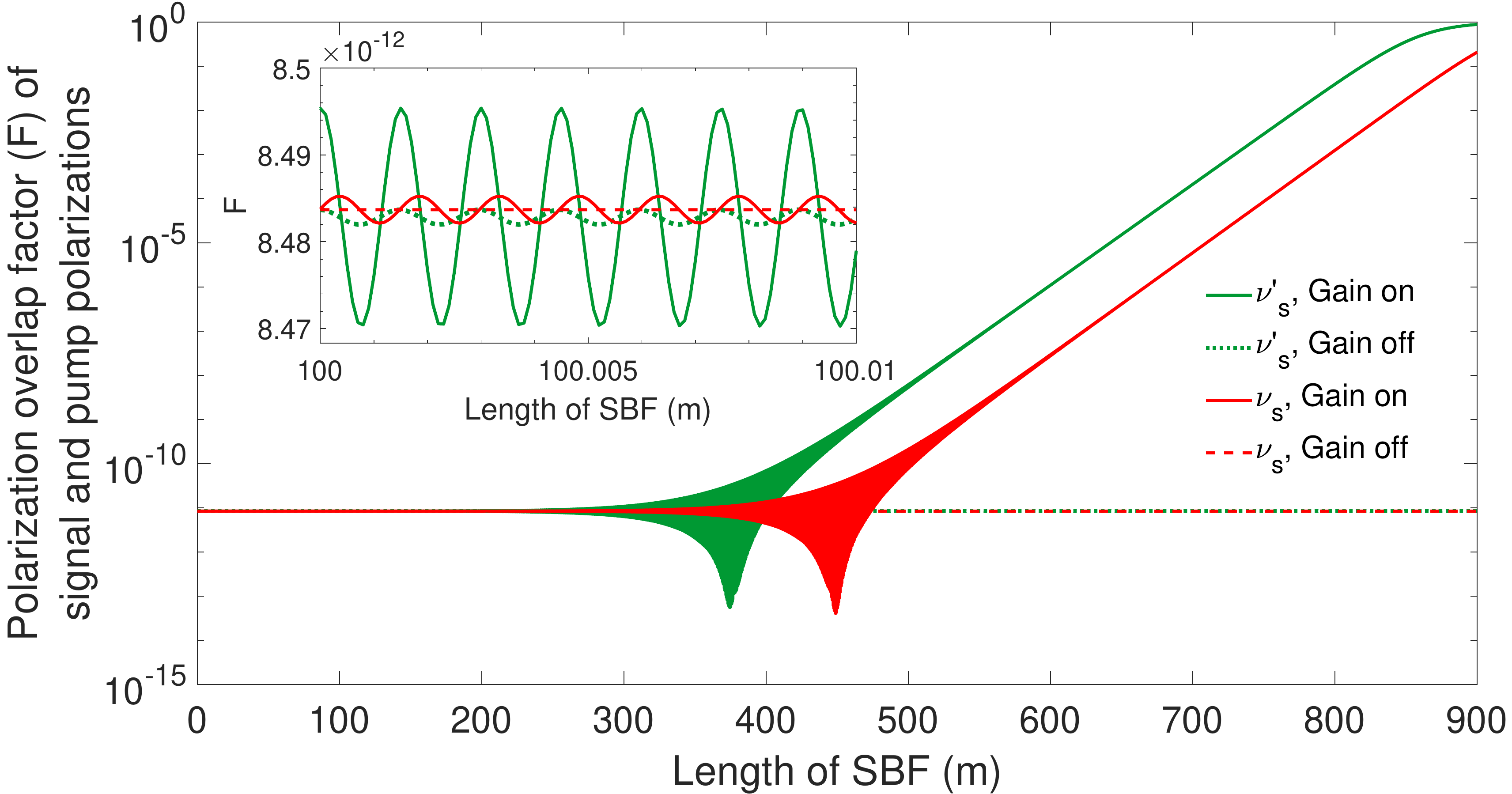}
\caption*{(c)}
\label{fig:deviation from signal eigenmode}
\end{subfigure}
\end{figure}
\begin{figure}[H]
    \centering
    \ContinuedFloat
\begin{subfigure}{0.6\textwidth}
\centering
\includegraphics[width=\textwidth]{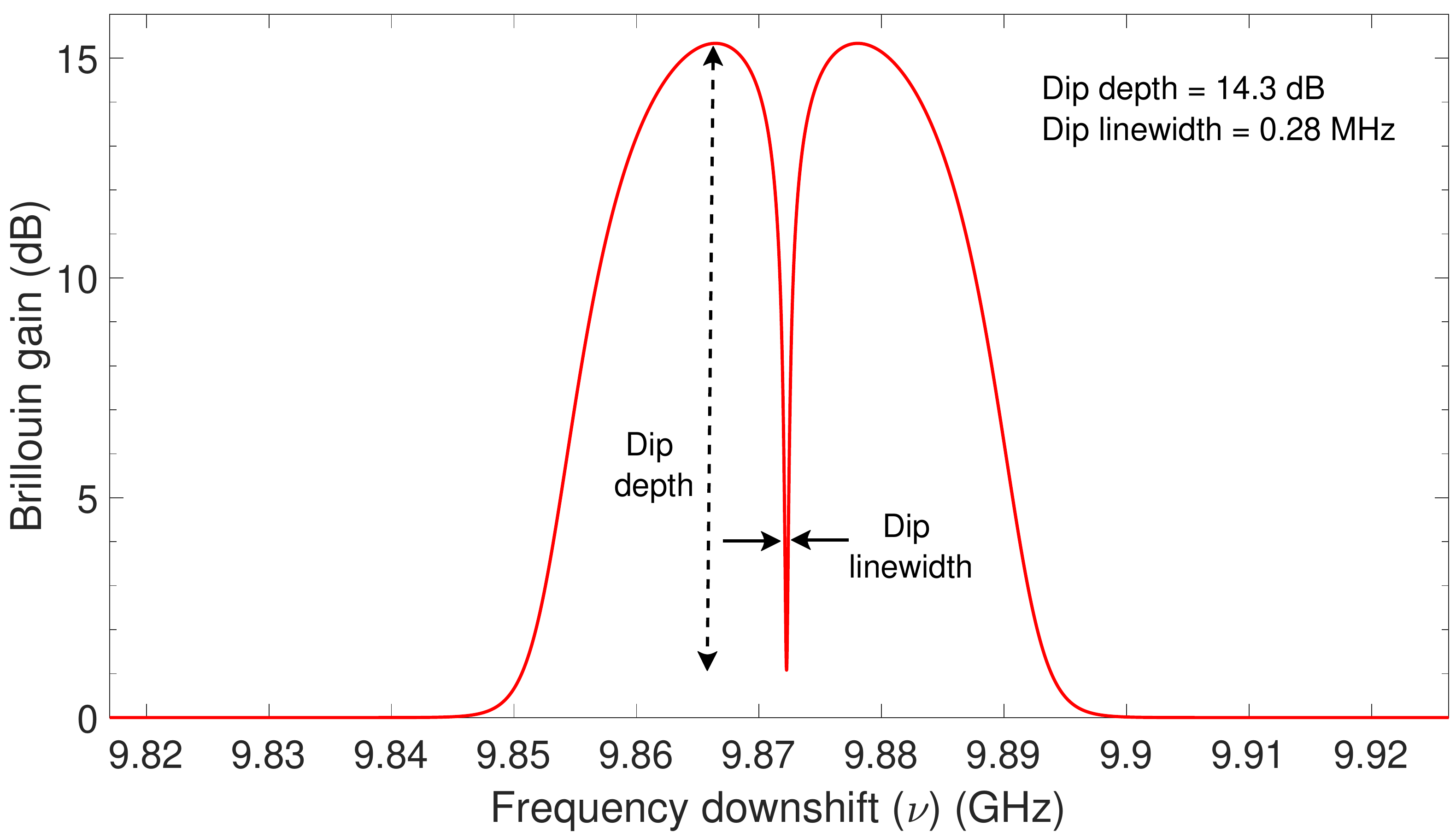}
\caption*{(d)}
\label{fig:simulation result}
\end{subfigure}
   \caption{(a) Simulated signal polarization evolution (indicated by the red and green traces with arrows on a Poincar\'e sphere) along the SBF, for signal frequencies $\nu_s = \nu_p$ - $\nu_B$ and (b) $\nu'_s = \nu_s + 0.5$ MHz. For both (a) and (b), the signal input polarizations corresponds to the eigenmodes $\boldsymbol{S_1}(\nu_s)$, which is an eigenmode for $\nu_s$, but not for $\nu'_s$. (c) Polarization overlap factor ($F$) of signal and pump polarization states as a function of the length of SBF for input signal polarization $\boldsymbol{S_1}(\nu_s)$ launched at frequencies $\nu_s$ and $\nu'_s$ with gain on ($P_{pump}$ = 14.75 dBm) and off. The inset is a zoomed in version near $z$ = 100 m, showing slight polarization oscillation of a non-eigenmode. (d) Simulation result demonstrating a spectral dip in SBS gain spectrum of SBF. The gain is plotted as a function of the frequency downshift ($\nu$, $\nu$ = $\nu_p-\nu_s$). $\nu_p$ is fixed at 192.97 THz, and the input signal polarization is fixed at $\boldsymbol{S_1}(\nu_p-\nu_B)$. The linewidth of the dip is 0.28 MHz when measured at 3 dB from the minimum, and the dip depth is 14.3 dB. $L_b(\nu_p)$ = 3 mm for (a) and (b) and 26 mm for (c) and (d). For all the simulations, pump polarization is aligned to $\boldsymbol{P_2}(\nu_p)$, $L_t$ = 3 mm, $P_{sig} = -$12 dBm, $\nu_B$ = 9.8723 GHz, and fiber loss is not considered.}
    \label{fig:Polarization pulling}
\end{figure}

Thus, the frequency dependence of the eigenmodes along with the polarization pulling effect of the Brillouin gain leads to an ultra-narrow spectral dip in the Brillouin gain spectrum (Fig. \ref{fig:Polarization pulling}d). A spectral dip with a linewidth of 0.28 MHz (measured at 3 dB from the minimum) is clearly observed. Here, we follow the conventional way of plotting the SBS gain as a function of frequency downshift ($\nu$), though the spectral dip is obtained in the optical domain, at the signal frequency ($\nu_s$).

To gain more insights into the behaviour of this spectral dip and how it is affected by linear birefringence and twist rate, let us define a Birefringence-to-Twist Ratio (BTR = $k_{bp}$/$k_t$ = $L_t$/$L_b(\nu_p)$). We first analyze the dip depth and linewidth as a function of beat length, at a fixed BTR of 1, meaning that $L_t$ is equal to $L_b$. As seen in Fig. \ref{fig:dip linewidth and depth for ratio 1}, the dip disappears if the beat length (and twist period) becomes too large. In the extreme, the spun fiber becomes an SMF. In the opposite direction, when the beat length (and twist period) decreases, the dip sharpens and deepens, until the dip depth and linewidth converge to limiting values. In other words, for a fixed BTR, when the beat length (and twist period) is sufficiently small, the dip reaches its sharpest limit and no longer narrows or deepens. This is useful for practical considerations. It means that one does not need to fabricate SBFs with impractically high birefringence or twist rate to achieve the narrowest dip. A beat length and twist period of a few mm is practically achievable.
\begin{figure}[H]
\centering
\begin{subfigure}{0.7\textwidth}
\centering
\includegraphics[width=\textwidth]{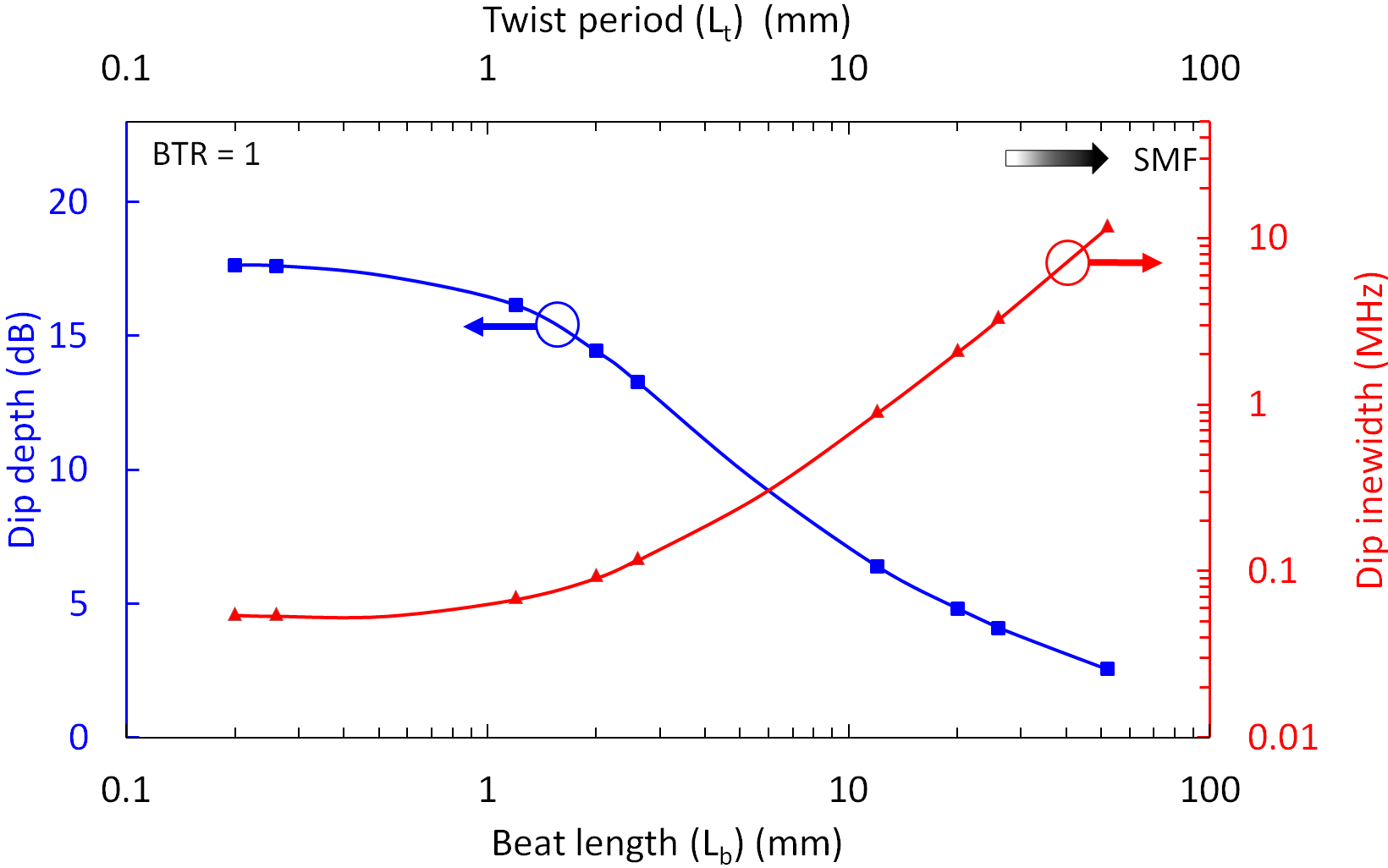}
\caption{}
\label{fig:dip linewidth and depth for ratio 1}
\end{subfigure}
\end{figure}
\begin{figure}[H]
\ContinuedFloat
\centering
\begin{subfigure}{0.7\textwidth}
\centering
\includegraphics[width=\textwidth]{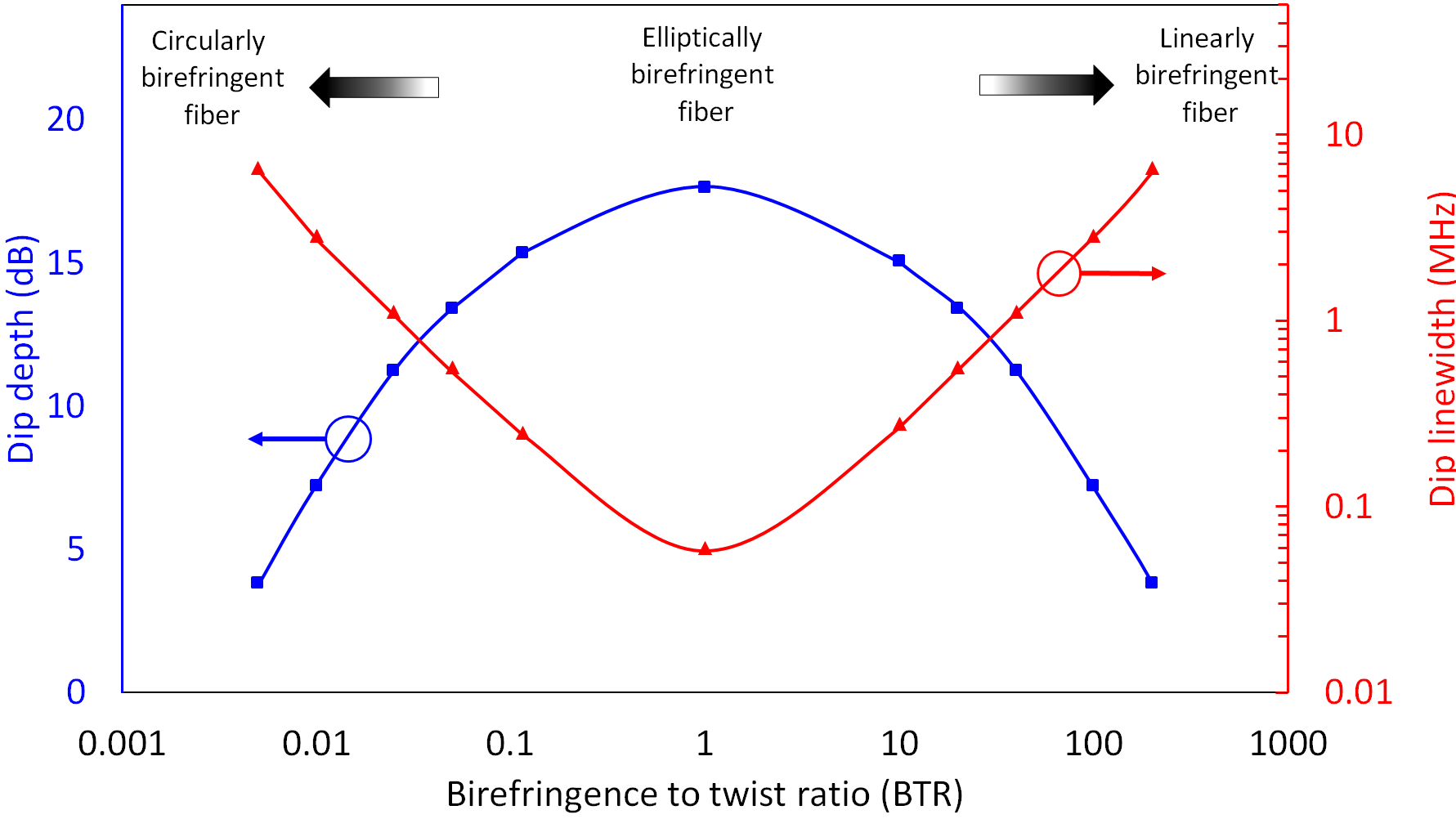}
\caption{}
    \label{fig:dip linewidth and depth}
\end{subfigure}
   \caption{Simulation results showing the depth and the linewidth of the spectral dip as a function of (a) $L_b (\nu_p)$ and $L_t$, when the BTR is fixed at 1, and (b) BTR, for a fixed $L_b(\nu_p)$ = 0.26 mm. For BTR = 1, the dip is most prominent i.e., linewidth is narrowest, and depth is highest. The simulation parameters are: $L$ = 900 m, $\nu_p$ = 192.97 THz, $P_{sig} = -$12 dBm, and $P_{pump}$ = 14.75 dBm. The symbols are simulation data points, and the lines are guides to the eye. Fiber loss is not considered for these simulations.}
    \label{fig:twist period beat length and BTR variation}
\end{figure}
Next, we analyze the dip behaviour as a function of BTR (Fig. \ref{fig:dip linewidth and depth}), keeping the beat length sufficiently small ($L_b$ = 0.26 mm). At very low BTRs (e.g., $<10^{-3}$), the SBF behaves like a circularly birefringent fiber, and for high BTRs (e.g., $> 10^3$), the SBF behaves like a linearly birefringent fiber (such as the PMF). At both extremes, the polarization eigenmodes become independent of frequency, and the dip disappears. On the other hand, when BTR lies in the range of [0.05, 20], the dip linewidth is <1 MHz, and the depth is high (>10 dB). Predictably, the dip is narrowest (with a linewidth of $\sim$ 0.06 MHz) and deepest (with a depth of 17.6 dB) at BTR = 1. This dip linewidth is equivalent to a Q-factor greater than 3 billion if a resonator was used. 

Through these analyses, we deduce that this spectral dip in the SBS gain spectrum is unique to elliptically birefringent media (such as SBF). It is not observed when both beat length and twist period become large (Fig. \ref{fig:dip linewidth and depth for ratio 1}), and it also disappears when either linear birefringence or circular birefringence dominates (Fig. \ref{fig:dip linewidth and depth}). This is why this dip has never been observed in SMF or PMF. This work is the first time such a spectral dip is ever analyzed and observed.
\par
Furthermore, the spectral location of this dip can be tuned in real-time (on demand) by changing either the pump frequency or the input polarization, the latter is indicated in Fig. \ref{fig:spectral position}. The linewidth and the depth of the dip can also be tuned, though not independently, by varying the pump power. Under a certain pump power, as we increase the pump power, the dip linewidth narrows and the depth deepens (Fig. \ref{fig:pump power}). This behaviour is analogous to the lasing phenomenon where the linewidth of a laser decreases due to the positive feedback received from the gain medium. Beyond a certain pump power, the polarization pulling becomes more significant with an increasing pump power, even for the eigenmodes $\boldsymbol{S_1}$, since it is not strictly orthogonal to $\boldsymbol{P_2}$ (also see Fig. \ref{fig:Polarization pulling}c). This causes the gain at the dip to increase with increasing pump power, and eventually, the dip disappears (Fig. \ref{fig:pump power}).
\begin{figure}[H]
\centering
\begin{subfigure}{0.65\textwidth}
\includegraphics[width=\textwidth]{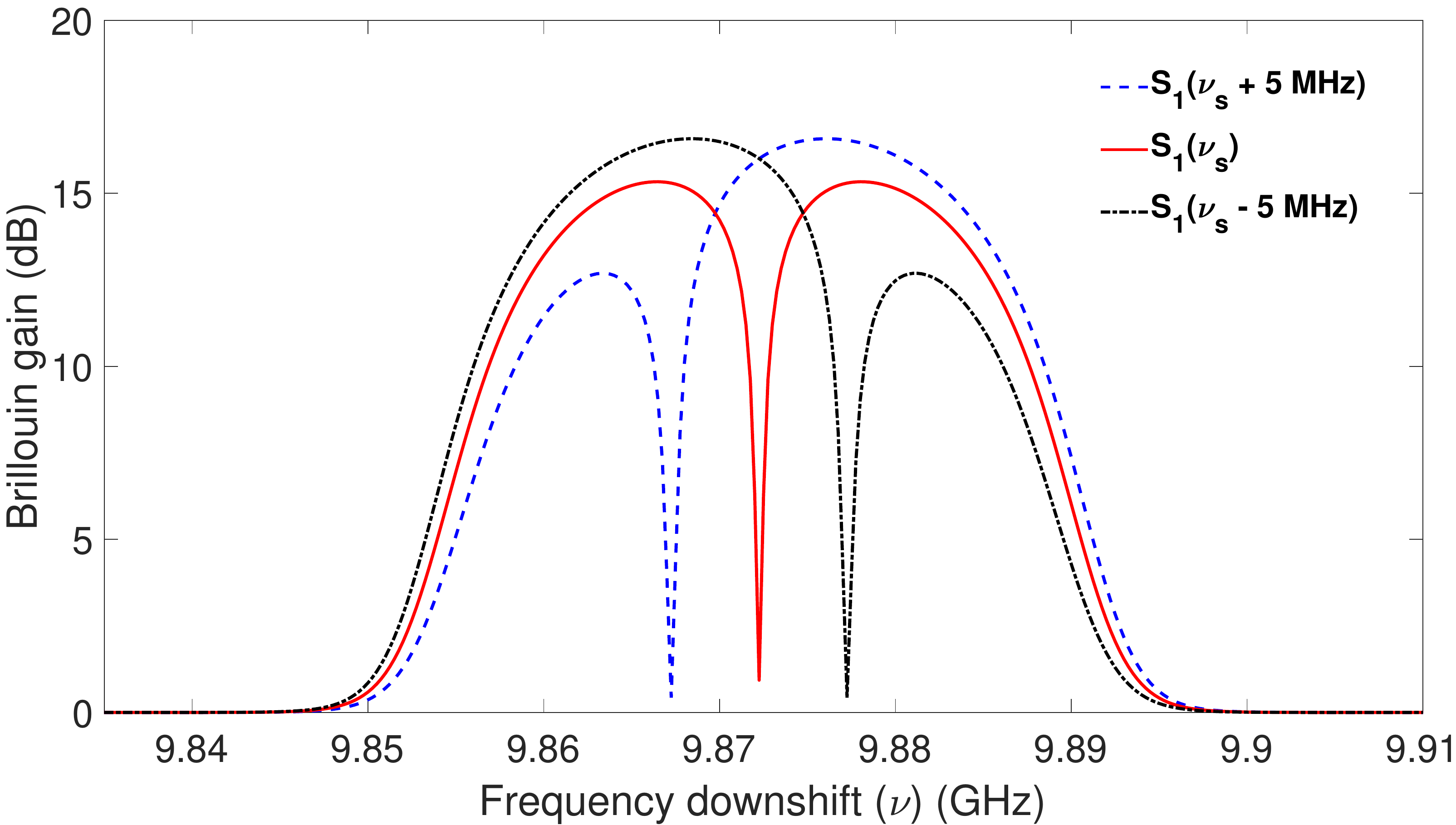}
\caption{}
\label{fig:spectral position}
\end{subfigure}
\end{figure}
\begin{figure}[H]
\ContinuedFloat
\centering
\begin{subfigure}{0.6\textwidth}
 \includegraphics[width=\textwidth]{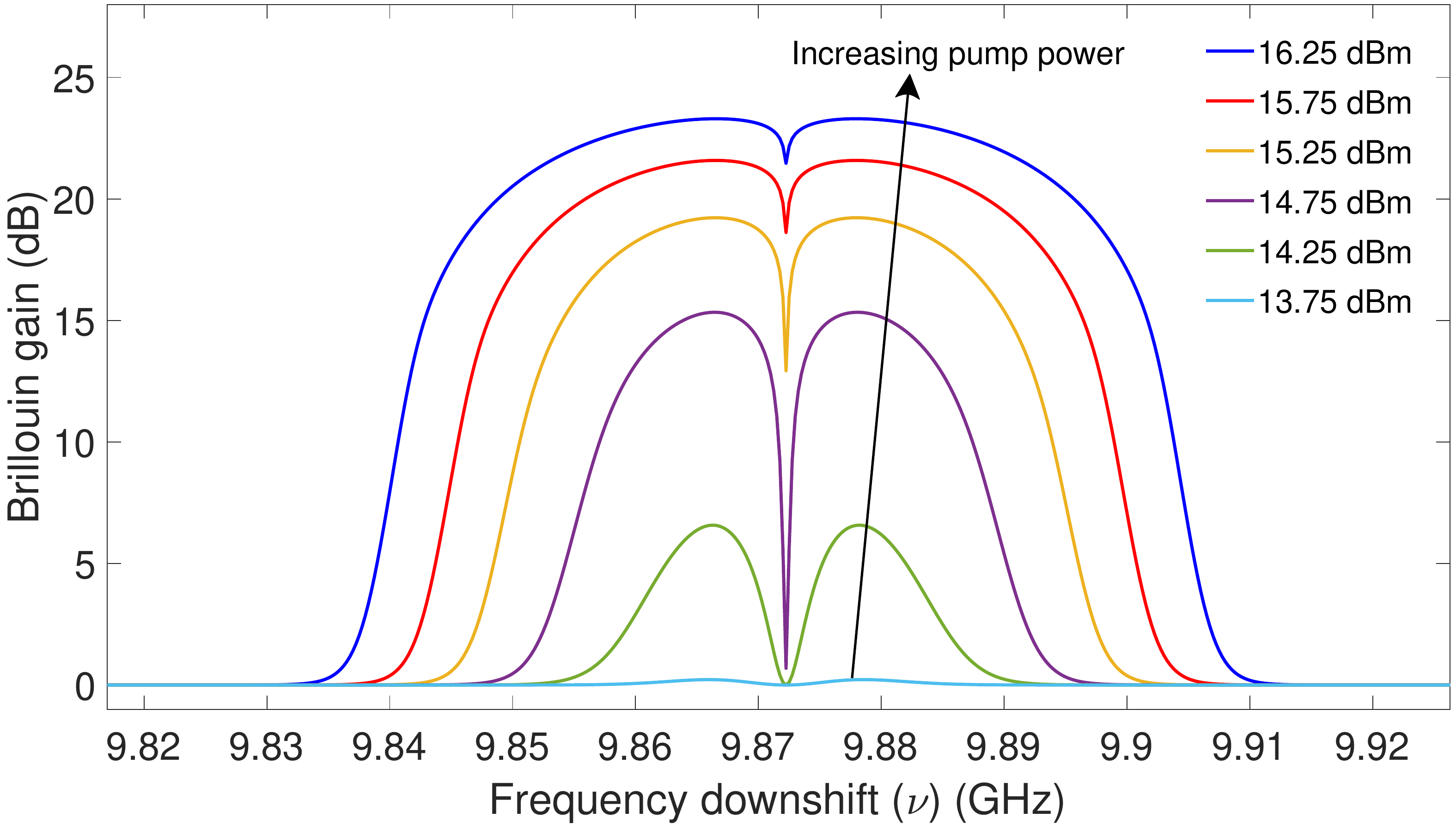}
 \caption{}
 \label{fig:pump power}
\end{subfigure}
   \caption{Simulation results showing the tunability of (a) the spectral position of the dip by varying the input signal polarization, while keeping the pump polarization and $P_{pump}$ (14.75 dBm) fixed; (b) the linewidth and the depth of the dip in the Brillouin gain spectrum of SBF by varying $P_{pump}$ for a fixed input signal polarization $\boldsymbol{S_1}(\nu_p-\nu_B)$. For both (a) and (b), $\nu_p$ = 192.97 THz, input pump polarization is aligned to $\boldsymbol{P_2}(\nu_p)$, $L$ = 900 m, $L_t$ = 3 mm, $L_b(\nu_p)$ = 26 mm, $P_{sig} = -$12 dBm, and $\nu_B =  9.8723$ GHz. Fiber loss is not considered for these simulations.}
\label{tunability of linewidth depth and location}    
\end{figure}
\par
In summary, by controlling the pump frequency, pump power and signal polarization, one has sufficient degrees of freedom to tune the dip frequency, linewidth, and depth, in real-time. This has an important practical significance in several applications. 
\section*{Experimental Results}
For an experimental demonstration, we use a commercial SBF manufactured by IVG Fiber \cite{IVG}. Our SBF has a beat length of 26 mm, twist period of 3 mm, and length of 900 m (see Fig. S7 in the supplementary to learn about the uniformity of our SBF). Note that this fiber does not have the optimal BTR of 1, but it is the only SBF available to us. The experimental setup (Fig. \ref{experimental arrangement}) splits a CW laser (wavelength = 1554.616 nm) into two paths. In the upper path, the pump field is amplified, and its polarization is adjusted before launching into SBF using a polarization controller. In the lower path, the signal field is amplified, and amplitude-modulated to produce two sidebands. The lower frequency sideband is selected by a tunable filter (AOS ultra-narrow filter with 0.12 nm passband bandwidth). The signal polarization is adjusted before launching into SBF, and its frequency is swept over the Brillouin gain range by sweeping the RF driving frequency of the electro-optic modulator (EOM).
\begin{figure}[H]
\centering
\begin{subfigure}{0.7\textwidth}
 \centering
 \includegraphics[width=\textwidth]{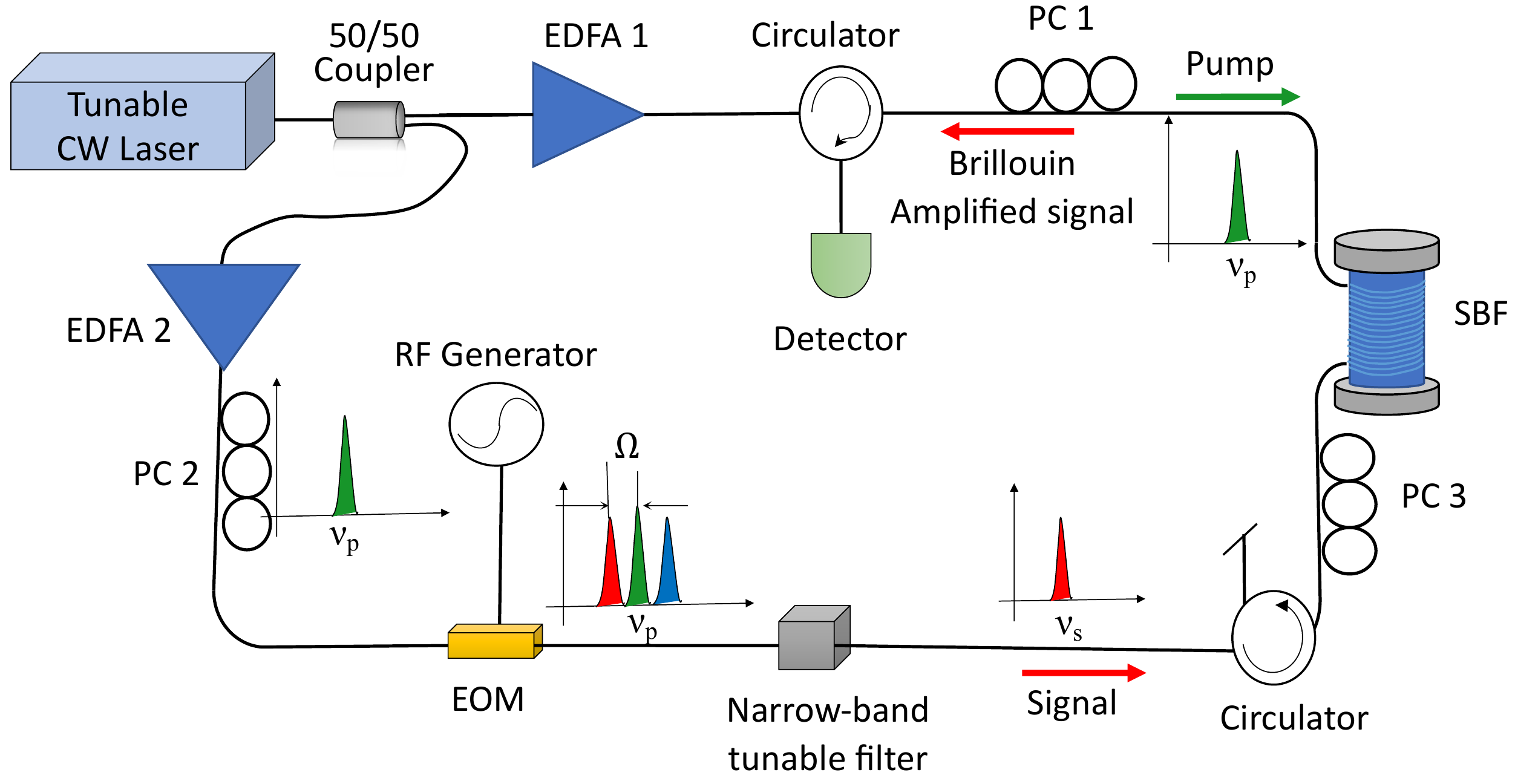}
    \caption{}
\label{experimental arrangement}
\end{subfigure}
\end{figure}
\begin{figure}[H]
    \centering
    \ContinuedFloat
\begin{subfigure}{0.7\textwidth}
 \centering
   \includegraphics[width=\textwidth]{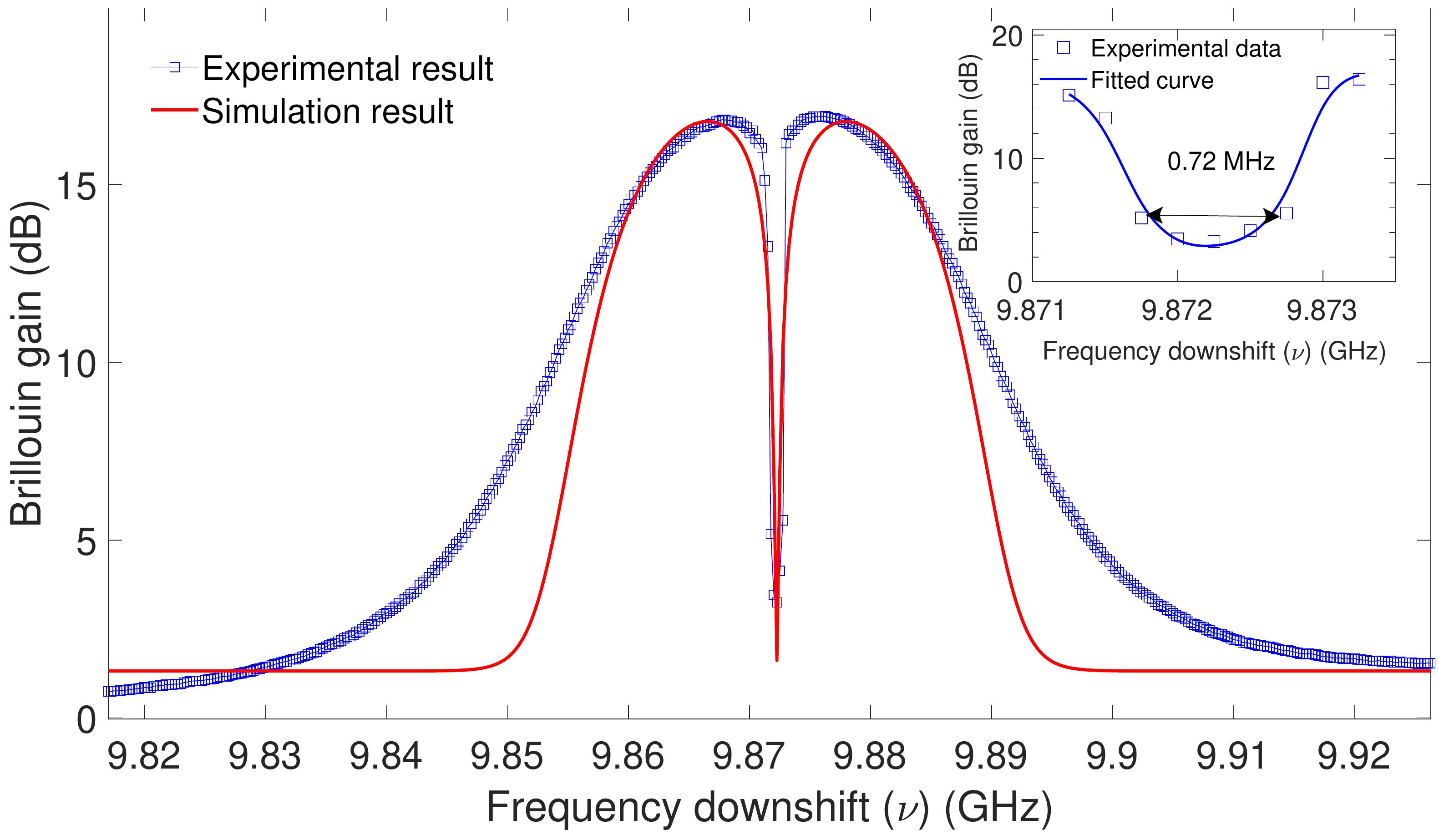}
   \caption{}
    \label{fig:Experimental simulation comparison}
\end{subfigure}
\caption{(a) Experimental setup to measure Brillouin gain: Signal frequency is downshifted from the pump by selecting the lower frequency sideband (red) of the modulated signal after the electro-optic modulator (EOM). Pump and signal polarizations are controlled by the polarization controllers PC1 and PC3, respectively; (b) Comparison of the SBS gain spectrum obtained via simulation and experiment. The simulation and experimental parameters are alike, and they are as following: $L$ = 900 m, $L_t$ = 3 mm, $L_b(\nu_p)$ = 26 mm, $P_{pump}$ = 16.9 dBm, $P_{sig} = -$12 dBm, pump wavelength = 1554.616 nm, and fiber loss = 5 dB/km. For the simulation result, we included a constant ASE noise in the pump to emulate the experimental conditions. This results in a constant background offset of the Brillouin gain. The inset shows a zoomed-in version of the experimental result near the dip.}
\end{figure}
\squeezeup

Due to the high sensitivity of the spectral dip to signal and pump polarization variation, we employ passive polarization stabilization in our experiment. We place most of the components (excluding the laser, amplifiers, spectral filter, RF modulator, and power meter) in a passive plexiglass enclosure. This prevents air flow through the arrangement, and makes it less prone to environmental disturbances. The room temperature variation is kept within 1$^{\circ}$ C. We observed that, with such passive stabilization, the dip is stable for the duration of measurement. 

The experimental and the simulation results are compared in Fig. \ref{fig:Experimental simulation comparison}, and good agreement between the two is evident. For the simulation result, we included a fiber loss of 5 dB/km, and a constant amplified spontaneous emission (ASE) noise in the pump field to emulate the experimental conditions. The ASE noise in pump results in a constant background offset of the Brillouin gain. The simulated dip linewidth is 0.28 MHz, whereas the experimental dip linewidth is 0.72 MHz, obtained from the least squares polynomial fit of the experimental data near the dip (see inset of Fig. \ref{fig:Experimental simulation comparison}). The experimental dip being wider than the simulation dip is likely due to the fact that a lower spectral resolution is used in the measurement (0.25 MHz) than in the simulation (0.01 MHz). The experimental gain shape on the wings of the Brillouin gain deviates from the simulation shape likely due to the following factors: (a) we have not accounted for the nonlinear polarization rotation in our model; (b) we have not accounted for the ASE noise in the signal; (c) our model assumes single, pure polarization input, whereas this is not true in practice, for the pump as well as for the signal. Though, these factors are only hypotheses at this stage and need to be rigorously investigated in the future.

Despite the discrepancy at the "wings" of the spectral gain shape, the simulation captures the ultra-narrow dip feature as well as its tunability, in high consistency with the experimental measurements, thus validating the mechanism we presented here as the cause for the ultra-narrow dip.
\begin{figure}[H]
\centering
\begin{subfigure}{0.63\textwidth}
 \centering
 \includegraphics[width=\textwidth]{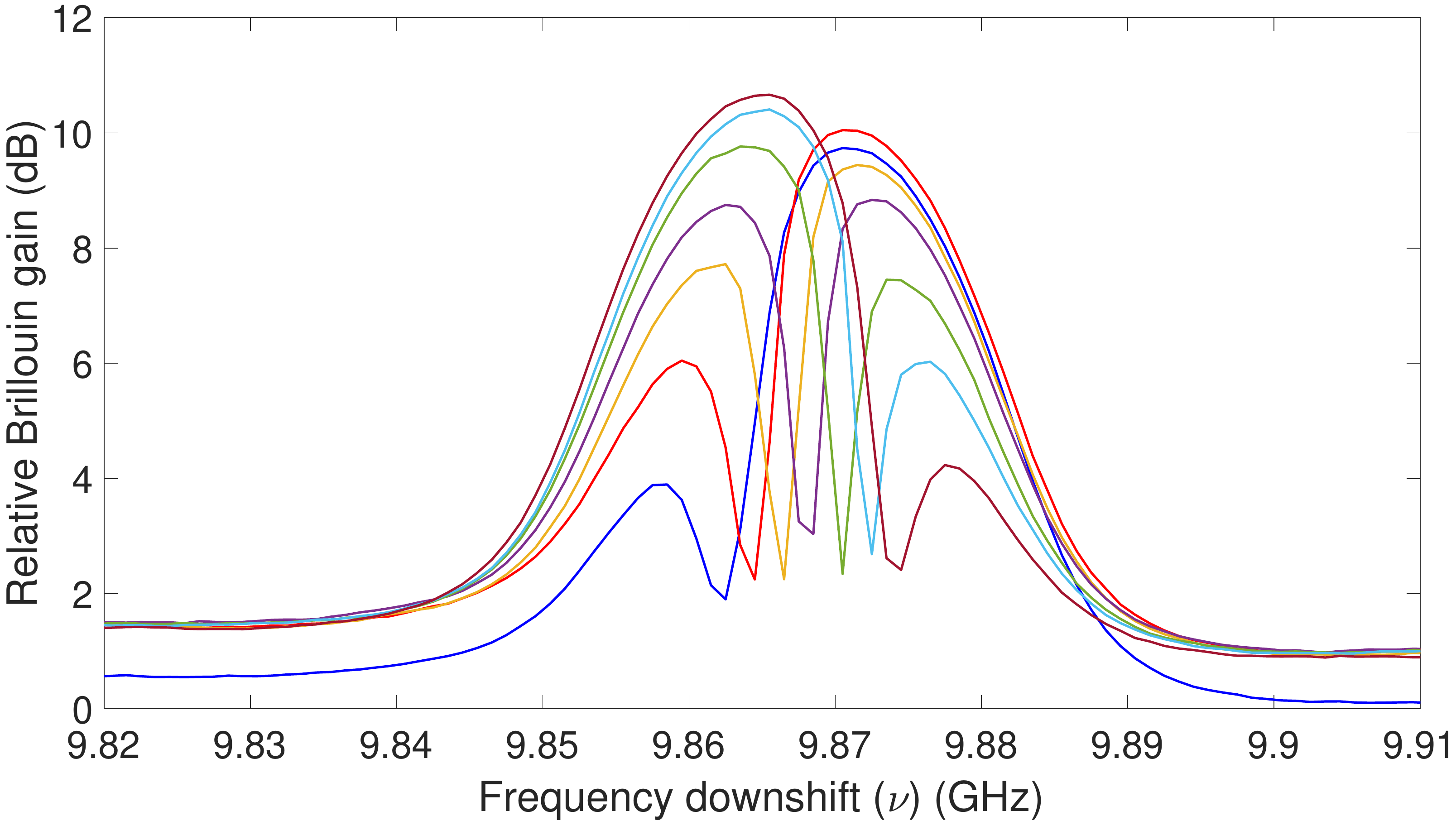}
 \caption{}
\label{fig:spectral position_experiment}
\end{subfigure}
\end{figure}
\begin{figure}[H]
    \centering
    \ContinuedFloat
\begin{subfigure}{0.65\textwidth}
\includegraphics[width=\textwidth]{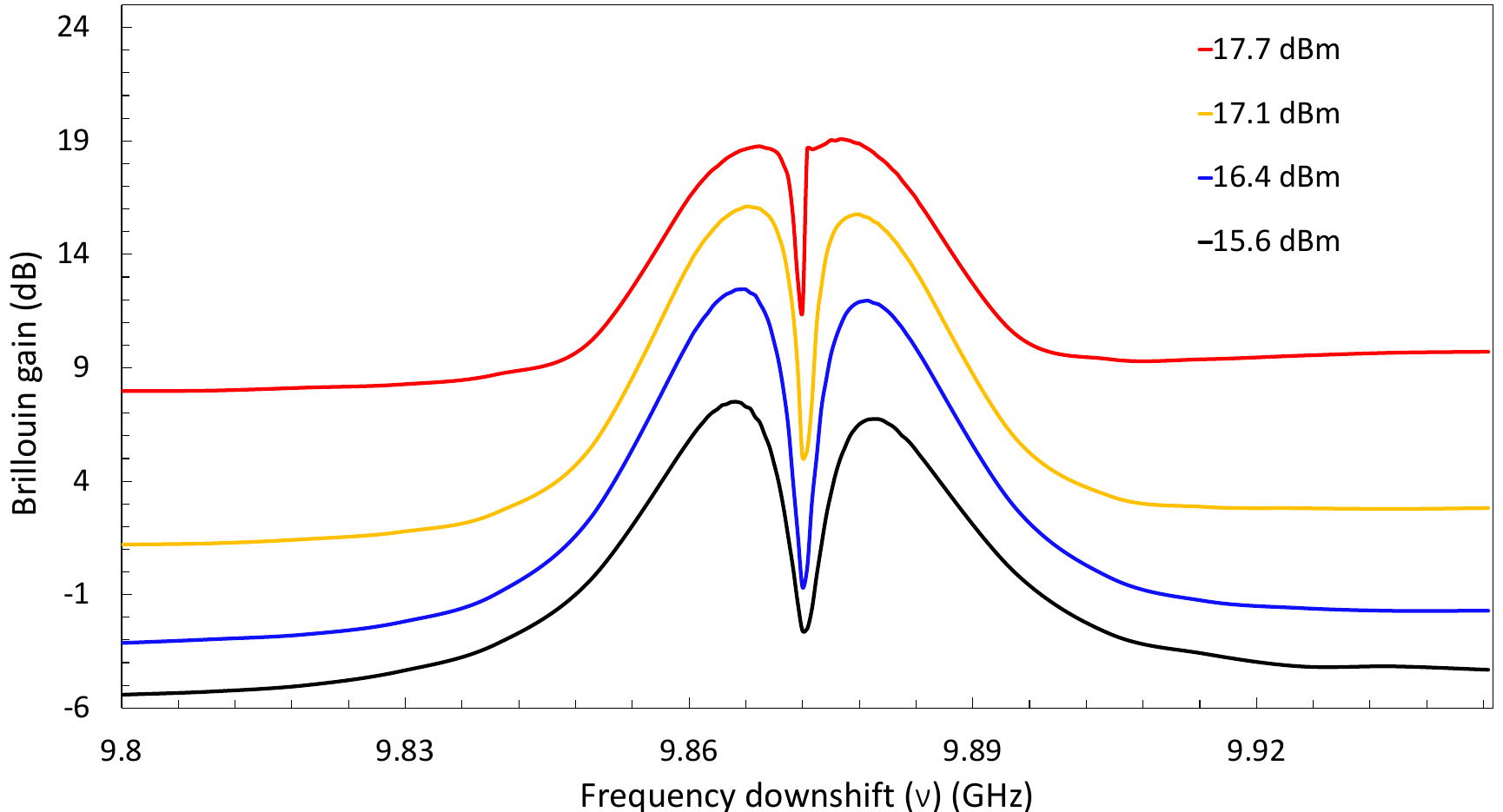}
 \caption{}
 \label{fig:pump power variation_experiment}
\end{subfigure}
   \caption{Experimental demonstration of (a) the tunability of the spectral position of the dip by varying the input signal polarization, while keeping the pump polarization, $P_{pump}$ (16.9 dBm), and $P_{sig}$ ($-$12 dBm) constant; (b) the tunability of the linewidth and the depth of the dip by varying $P_{pump}$, while keeping the input polarizations and $P_{sig}$ constant ($-$16.2 dBm). The fiber parameters are as following: $L$ = 900 m, $L_t$ = 3 mm, and $L_b(\nu_p)$ = 26 mm.}
   \label{experiment tunability}
\end{figure}
The tunability of the spectral location of this dip by changing the input polarization, and the tunability of the linewidth and the depth of dip by varying input pump power are experimentally verified (see Fig. \ref{fig:spectral position_experiment} and Fig. \ref{fig:pump power variation_experiment}). Note, for low pump powers, the gain becomes negative on the "wings" due to fiber loss, while for high pump powers, the gain has a positive background on the "wings", due to ASE noise in the pump and the signal. The ASE-induced background can be largely avoided if one uses a high power laser without amplifiers.

The high sensitivity of the spectral location of the dip to variation in input signal polarization, both predicted by our model (Fig. \ref{fig:spectral position}) and observed experimentally (see Fig. \ref{fig:spectral position_experiment} and Fig. S6 in the supplementary), further validates the dip forming mechanism provided in this work, and excludes the possibility of spectral hole burning \cite{takushima1995spectral,kovalev2000observation,kovalev2002waveguide,stepien2002origin} being the mechanism behind the dip (also see Fig. S6 in the supplementary material where we compare the SBS gain for co-polarized and orthogonally polarized inputs). Such high polarization-sensitivity cannot be explained by spectral hole burning.

\section*{Discussion}
As we have seen, both theoretically and experimentally, the sub-MHz spectral dip is a result of polarization pulling in an elliptically birefringent medium. The filtering effect in a birefringent medium is reminiscent of birefringent filters, of which Lyot filter \cite{evans1949birefringent} and \v{S}olc \cite{vsolc1965birefringent,evans1949birefringent} filter are two well-known examples. The SBF is analogous to a fan-\v{S}olc filter \cite{vsolc1965birefringent,evans1949birefringent} except for three facts: (a) the birefringence axis is continuously rotating in SBF, whereas in a fan-\v{S}olc filter, discrete birefringent plates have discrete rotation angles; (b) the overall rotation angle in SBF is not limited to $\pi$/2, and in fact is many orders of magnitude larger; (c) there are no polarizers in SBF as in the \v{S}olc filter. It is this last point that warrants further discussion. In our demonstration, instead of polarizers (which in essence introduce polarization-dependent loss), we have Brillouin gain, which introduces polarization-dependent gain. This is a crucial distinction, and is the reason why we can achieve such a narrow spectral feature. In fact it can be shown that when the same length of the SBF (that is, the same amount of polarization rotation) is placed in between two polarizers, the resulting \v{S}olc filter feature is on the order of tens of GHz (see supplementary material Fig. S5). So, why are we able to obtain sub-MHz features  when we use polarization-dependent gain? The answer lies in the positive feedback the gain produces. Much like the linewidth narrowing when gain builds up in a lasing process and the gain narrowing in the SBS process \cite{boyd1990noise}, the dip becomes narrower and narrower when gain increases (see Fig. \ref{fig:pump power} and Fig. \ref{fig:pump power variation_experiment}), resulting in a 5 orders-of-magnitude narrowing of the passive filter feature (see supplementary material Fig. S5). 

In a nutshell, the innovation of our method lies in the mechanism of enhancing birefringent filtering using a polarization-dependent gain to create a positive feedback. It is to our knowledge the first proposal and demonstration of such a mechanism, which can be broadly implemented using other elliptically birefringent media and gain mechanisms, and therefore can potentially lead to a paradigm shift in the pursuit of ultra-narrow spectral features and their applications.

For example, the strong polarization pulling effect of SBS in the vicinity of the spectral dip leads to a high dispersion within the narrow spectral region, which can be used for realizing group delays and information storage applications (see Fig. S1 in the supplementary material for the theoretical predictions of the phase response and the delay response associated with the spectral dip). Another application can be a narrow-band tunable spectral filter or microwave photonic filter which could be designed by combining the spectral dip in the gain medium with a broadband attenuator. Additionally, the high sensitivity of the spectral dip to polarization variations can be exploited for high-precision sensing of current, stress, and temperature.

Although the high sensitivity of the spectral dip to input polarization variations can enable ultrahigh sensitive measurements, it can be a drawback for some other applications, as well. For example, it is necessary to precisely align the signal and pump polarizations to the polarization eigenmodes of SBF to generate the spectral dip. Also, as we have shown in Figs. \ref{fig:pump power} and 6b, there is a limited range of gain for which this dip can be obtained. The dip will disappear for very high or very low gain. This can lead to performance limitations in some of the above-mentioned applications. Furthermore, the maximum depth of the spectral dip in our experiments is limited to $\sim$ 14 dB due to our fiber BTR and input parameters. A higher dip depth can be achieved by choosing optimal input powers and an SBF with optimal BTR of 1 (see Fig S8 in the supplementary material where simulation shows that a dip depth of 37 dB can be obtained for an optimal BTR of 1, $P_{pump}$ = 14.75 dBm, and $P_{sig}$ = $-$35 dBm). Other SBS-based approaches such as \cite{yelikar2020analogue} offer an extinction ratio of 20-40 dB. 

In summary, we have demonstrated a resonator-free approach of generating a sub-MHz tunable spectral dip at room-temperature by exploiting polarization pulling in a medium with frequency-dependent polarization eigenmodes. As a specific realization, we have experimentally demonstrated a 0.72 MHz spectral dip in the Brillouin gain spectrum of a commercial spun fiber. The observed dip linewidth of 0.72 MHz is equivalent to a Q factor of $\sim$ 267 million if a resonator were used. The linewidth, depth and the location of the dip can be tuned on demand by controlling the pump frequency, the pump power, and the input polarization of the signal. Moreover, with an optimal spun birefringent fiber, the dip linewidth is predicted to be as low as $\leq$ 0.1 MHz, corresponding to a Q factor of 2 billion. 

Even though the current analysis of this spectral dip is carried out in fiber, this ultra-narrow feature can potentially be realized in integrated waveguides, using, for instance, chiral birefringent material or a helical waveguide. The gain mechanism is also not limited to SBS, and one can make use of polarization-dependent gain such as Raman gain or parametric gain. The essential elements required to realize such narrow spectral dips are a polarization-dependent gain and a rotating birefringence with frequency-dependent polarization eigenmodes.

The simplicity in the implementation of this technique, as well as the ultra-narrow spectral feature and the easily attainable tunability of the dip, may open a wide range of potential applications, such as ultrahigh resolution optical sensing, ultra-narrow band tunable optical filters for microwave photonics, fast-light and information storage applications.

The preliminary results of this research were presented at Conference on Lasers and Electro-optics \cite{choksi2021ultra} and Frontiers in Optics \cite{choksi2021resonator}.

\section*{Acknowledgements}
The authors gratefully acknowledge Natural Science and Engineering Research Council (NSERC) (RGPIN-2019-07019, RGPAS-2019-00113, and CREATE 484907-16 to L.Q.), Canada Foundation for Innovation (CFI) (Innovation Fund 33415, Leaders Opportunity Fund 203429, and New Opportunities Fund 9650 to L.Q.) for funding this research. Y.L.  acknowledges financial support from International Postdoctoral Exchange Fellowship sponsored by the China Postdoctoral Council and Wuhan University of Technology. 

\section*{Author Contributions}
L.Q. supervised the project. N.C. formulated and coded the theoretical framework, performed the calculations and data analysis, and together with L.Q. determined the underlying mechanism and significance of the dip. L.Q. conceived and designed the experiment. R.G. made the first experimental observation of the spectral dip. Y.L. improved the experimental data acquisition and along with R.G. showed the tunability of the dip, experimentally. N.C. and L.Q. experimentally determined that homogeneous gain broadening is the dominant broadening mechanism. N.C. and L.Q. wrote the paper and revised it. 

\section*{Competing Interests}
The authors declare no competing interests.

\section*{Data Availability}
The data that support the findings of this study are available in an open-access public repository \cite{Choksi2022}.

\section*{Code Availability}
The codes used for simulations are available from the corresponding authors upon reasonable request.

\bibliography{Manuscript}

\end{document}

\title{Supplementary information: Sub-MHz spectral dip in a resonator-free twisted gain medium}

\author{Neel Choksi\authormark{1*}, Yi Liu\authormark{1}, Rojina Ghasemi\authormark{1}, Li Qian \authormark{1*}} %and Author Three\authormark{2,3}}
\address{\authormark{1}Department of Electrical and Computer Engineering, University of Toronto, 10 Kings College Road, Toronto, Ontario M5S 3G4, Canada}
%\\
%\authormark{2}Publications Department, The Optical Society (OSA), 2010 Massachusetts Avenue NW, Washington, DC 20036, USA\\
%\authormark{3}Currently with the Department of Electronic Journals, The Optical Society (OSA), 2010 Massachusetts Avenue NW, Washington, DC 20036, USA}
%\email{\authormark{*}xyz}
\email{\authormark{*}neel.choksi@mail.utoronto.ca, l.qian@utoronto.ca} %% email address is required
\section{Phase Response and Delay Response Associated With Spectral Dip in SBS Gain}
The highly dispersive  nature of the ultra-narrow spectral dip leads to a sharp phase and delay response. The Kramers-Kronig relationship allows us to relate the phase response of a causal system to its amplitude response, if the amplitude response is known for all frequencies \cite{boyd2019nonlinear,ohta1988comparison}. If we assume that the stimulated Brillouin (SBS) gain with a spectral dip is the only amplitude response in a very large spectral region, then we can calculate the phase response associated with it. 

We take the following approach to get an approximate estimation of the group delay that such an ultra-narrow spectral dip can induce. We treat spun birefringent fiber (SBF) as a black box with an overall gain of $G(\omega)$, defined as the ratio of output to input signal power. We also assume that the gain and phase accumulations are uniform along the length of SBF (not strictly true due to pump depletion, but it gives us a good approximation). Then, we can obtain a gain parameter $g(\omega) (m^{-1})$ from $G(\omega)$ through:
\begin{align}
    G(\omega) &= exp(g(\omega)L)) \\
    g(\omega) &= \frac{ln(G(\omega))}{L}\label{Gain spectrum calculation},
\end{align}
where $\omega$ is the frequency of the signal ($\omega = 2\pi\nu_s$), and $L$ is the length of SBF. Next, we relate the phase response ($\phi(\omega)$) of SBS and $g(\omega)$ to the real ($n_r(\omega)$) and imaginary part ($n_i(\omega)$) of the complex refractive index ($ n(\omega) = n_r(\omega) + in_i(\omega)$), respectively:
\begin{align}
    exp\Big(\frac{i\omega n(\omega) L}{c}\Big) &= exp\Big(-\frac{\omega n_i(\omega) L}{c}\Big)exp\Big({\frac{i\omega n_r(\omega) L}{c}}\Big) = exp\Big({\frac{g(\omega)L}{2}}\Big)exp\Big({i\phi(\omega)}\Big) \label{refractive index and gain coefficient}\\
        n_i(\omega) &= \frac{-g(\omega)c}{2\omega} \label{imaginary part}\\
     \phi(\omega) &= \frac{\omega n_r(\omega)L}{c} \label{phase response}
\end{align}
Here, $c$ is the speed of light in vacuum. $n_r$ and $n_i$ are related to each other by the Kramers-Kronig \cite{boyd2019nonlinear} relationship shown below:
\begin{equation}
    n_r(\omega) = 1 + \frac{2}{\pi}P\int_{0}^{\infty}  \frac{\omega'n_i(\omega)}{\omega'^2 - \omega^2 }d\omega' \label{kramers kronig}
\end{equation}
Here, $P$ denotes that the principal value of the Cauchy integral should be considered. 

We solve the Kramers-Kronig integral (Eq. \eqref{kramers kronig}) using Maclaurin series \cite{ohta1988comparison}, and obtain $n_r$ from $n_i$, and then, we obtain the phase response ($\phi(\omega)$) for this interaction using Eq. \eqref{phase response}. The group delay ($\tau$) is related to $\phi(\omega)$ through the following relation:
\begin{equation}
    \tau = \frac{d\phi(\omega)}{d\omega}
\end{equation}
\par 
\begin{figure}[H]
   \centering
   \begin{subfigure}{0.65\textwidth}
        \includegraphics[width=\textwidth]{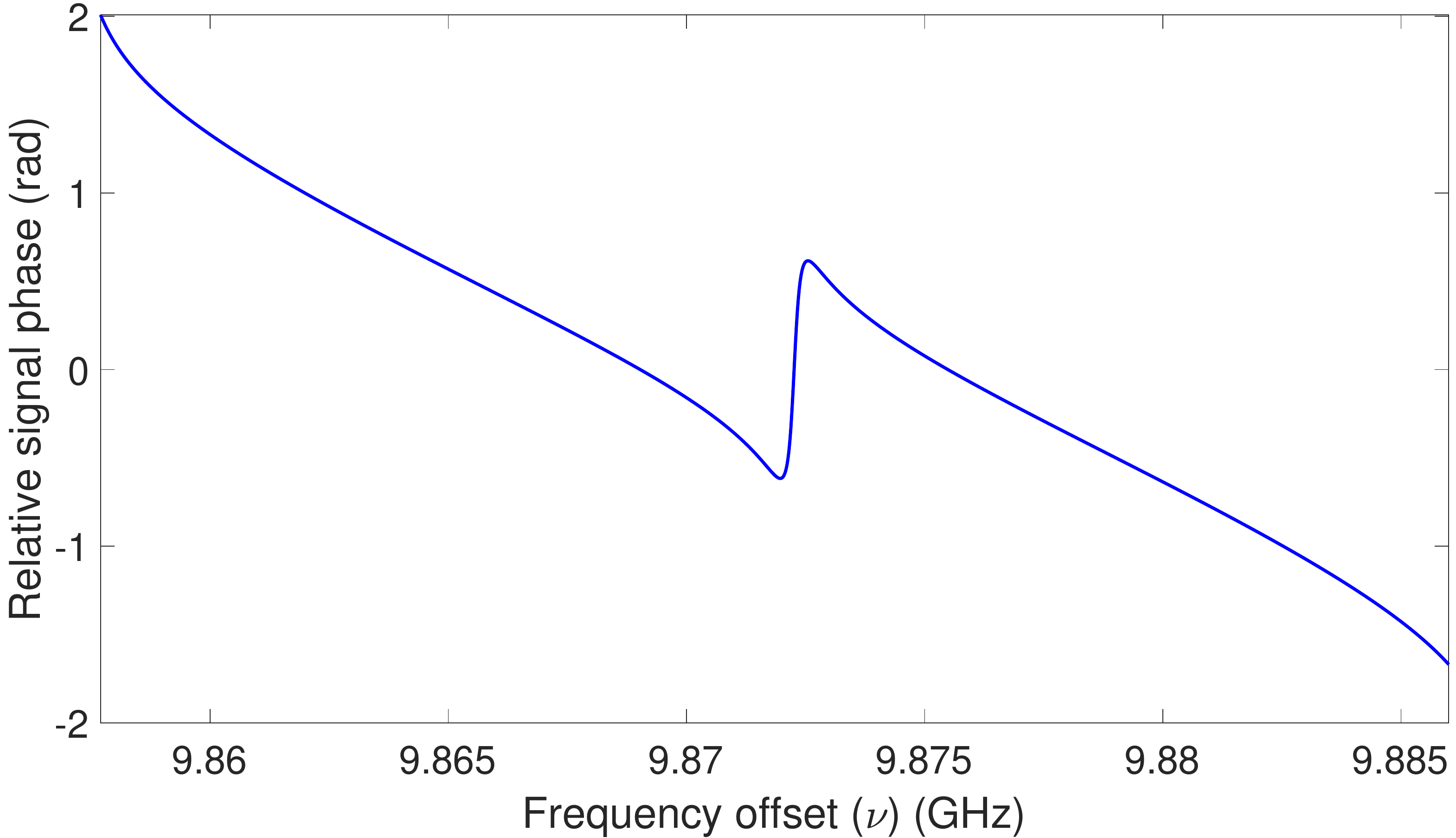}
        \centering
        \caption{}
        \label{fig:SBS phase response}
    \end{subfigure}
    \end{figure}
    \begin{figure}[H]
    \ContinuedFloat
    \centering
    \begin{subfigure}{0.65\textwidth}
        \centering
       \includegraphics[width=\textwidth]{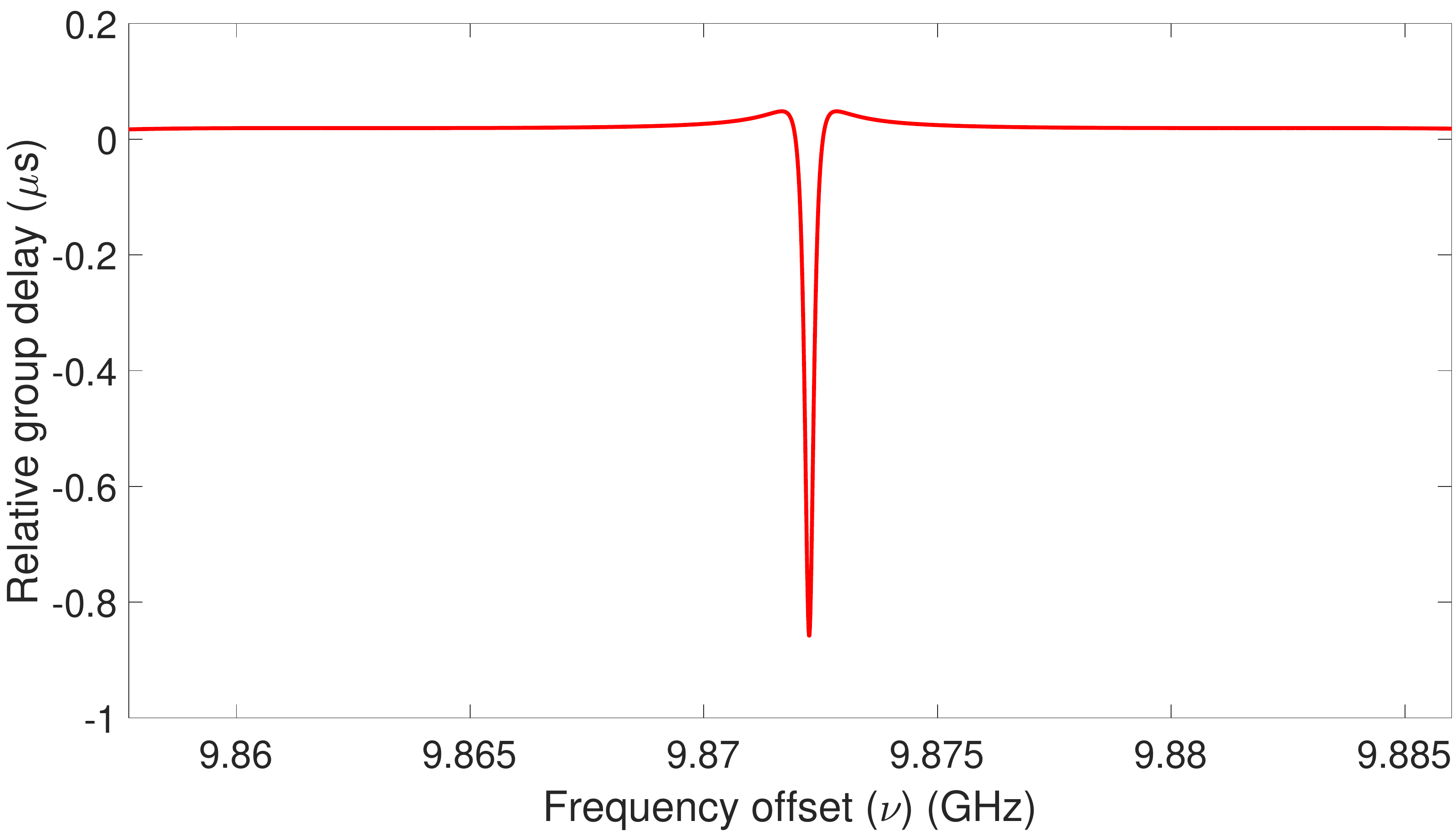}
        \caption{}
        \label{fig:SBS group delay}
    \end{subfigure}
    \caption{Theoretical prediction of (a) the phase response of SBS in SBF (b) group delay realized using SBS in SBF. The simulation parameters are: $L$ = 900 m, $L_t$ = 3 mm, $L_b(\nu_p)$ = 26 mm, $P_{pump}$ = 14.75 dBm, $P_{sig} = -$12 dBm, and $\nu_p$ = 192.97 THz. Fiber loss is not considered for these simulations.}
    \label{fig:Slow light}
\end{figure}
The phase and delay responses for the SBS gain spectrum (Fig. 2d in the main text) are plotted in Fig. \ref{fig:Slow light}. A relative group delay of $-0.858$ $\mu$s is predicted using our approach, which is comparable to the state-of-the-art SBS-based fast-light systems \cite{yelikar2020analogue}. Furthermore, this group delay can be tuned on-demand by changing input power, polarization, and frequency.
\section{Deviation from the Polarization Eigenmodes}
The polarization pulling effect is weak when both pump and signal are launched in their respective eigenmodes $\boldsymbol{P_2}$ and $\boldsymbol{S_1}$. Here, we examine what happens when the launch polarization of either pump or signal deviates from its eigenmode. We define three parameters $F$, the polarization overlap factor between signal and pump along the length of SBF (same as the definition in main text), $\delta_s$, the deviation of input signal polarization from the signal eigenmode $\boldsymbol{S_1}$, and $\delta_p$, the deviation of input pump polarization from the pump eigenmode $\boldsymbol{P_2}$. 
\begin{align}
    &F(z) = \Bigg|\frac{\Vec{A}^\dagger_{rs}(z).\Vec{A}_{rp}(z)}{|A_{rs}(z)||A_{rp}(z)|}\Bigg|^2 \label{overlap as a function of length}\\ 
    &\delta_s = 1 - \Bigg|\frac{\Vec{A}^\dagger_{rs}(0).\boldsymbol{S_1}(\nu_s)}{|A_{rs}(0)||\boldsymbol{S_1}|}\Bigg|^2 \label{overlap with signal eigenmode}\\
    &\delta_p = 1 - \Bigg|\frac{\Vec{A}^\dagger_{rp}(0).\boldsymbol{P_2}(\nu_p)}{|A_{rp}(0)||\boldsymbol{P_2}|}\Bigg|^2 \label{overlap with pump eigenmode}
\end{align}
\begin{figure}[H]
\centering
\includegraphics[scale=0.3]{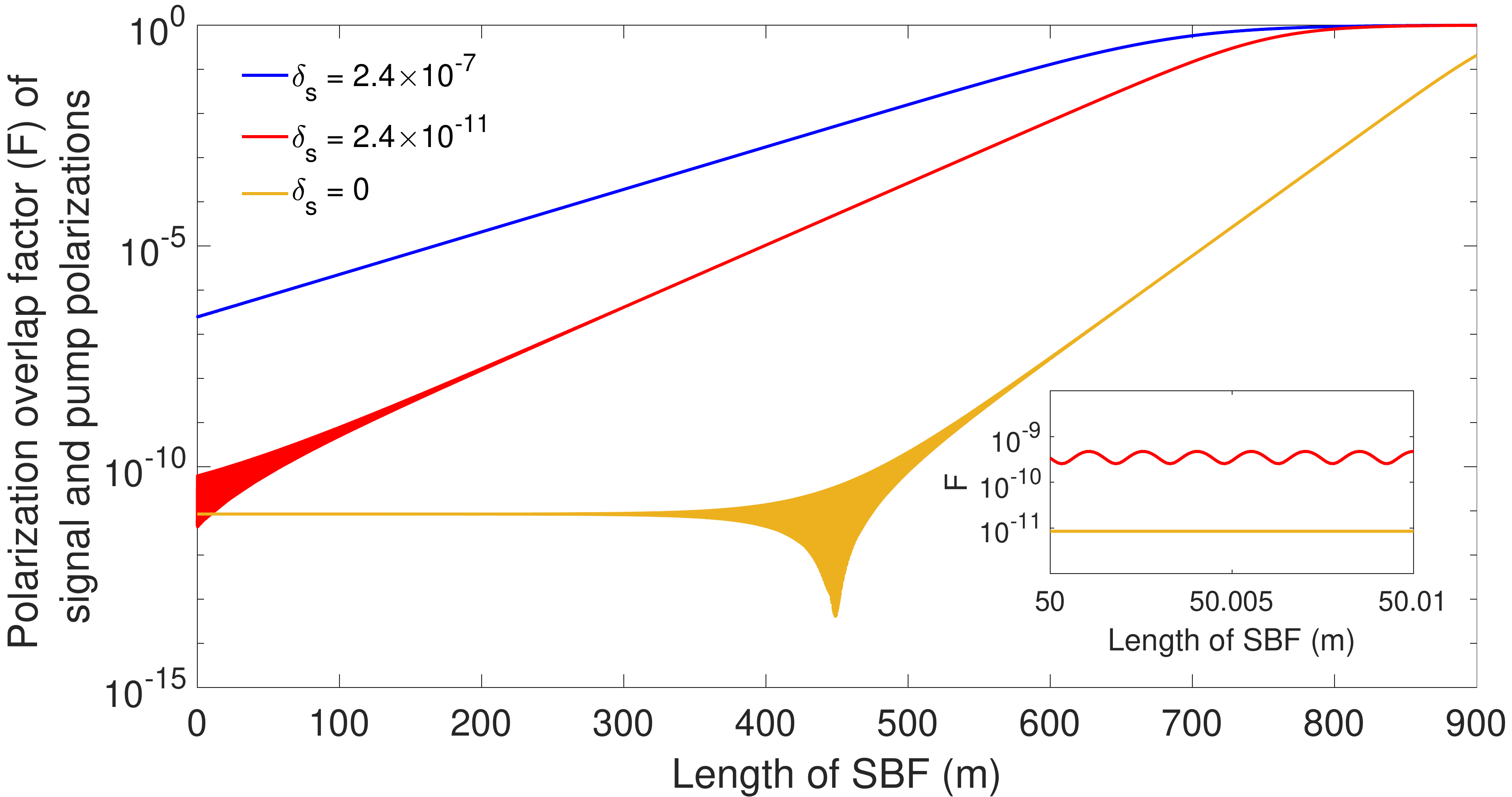}
\caption{Polarization overlap factor ($F$) of signal and pump polarization states as a function of the length of SBF for different $\delta_s$ from $\boldsymbol{S_1}(\nu_p-\nu_B)$. Refer to Eqs. \eqref{overlap as a function of length} and \eqref{overlap with signal eigenmode} for the definition of $\delta_s$ and $F$. The inset is a zoomed in version near $z$ = 50 m, showing slight polarization oscillation of a non-eigenmode. For all the cases, the input pump polarization is aligned to  $\boldsymbol{P_2}(\nu_p)$, $L_t$ = 3 mm, $L_b(\nu_p)$ = 26 mm, $\nu_p = 192.97$ THz, $\nu_B$ = 9.8723 GHz, $P_{sig} = -$12 dBm, and $P_{pump}$ = 14.75 dBm, and fiber loss is not considered}
\label{fig:deviation from signal eigenmode}
\end{figure}
\squeezeup
In Fig. \ref{fig:deviation from signal eigenmode}, we quantitatively show that for a small deviation from the signal eigenmode ($\delta_s$ = $2.4\times10^{-11}$), the signal field experiences a high polarization pulling effect of pump (see the red curve in Fig. \ref{fig:deviation from signal eigenmode}). 

\begin{figure}[H]
    \centering
    \includegraphics[scale=0.3]{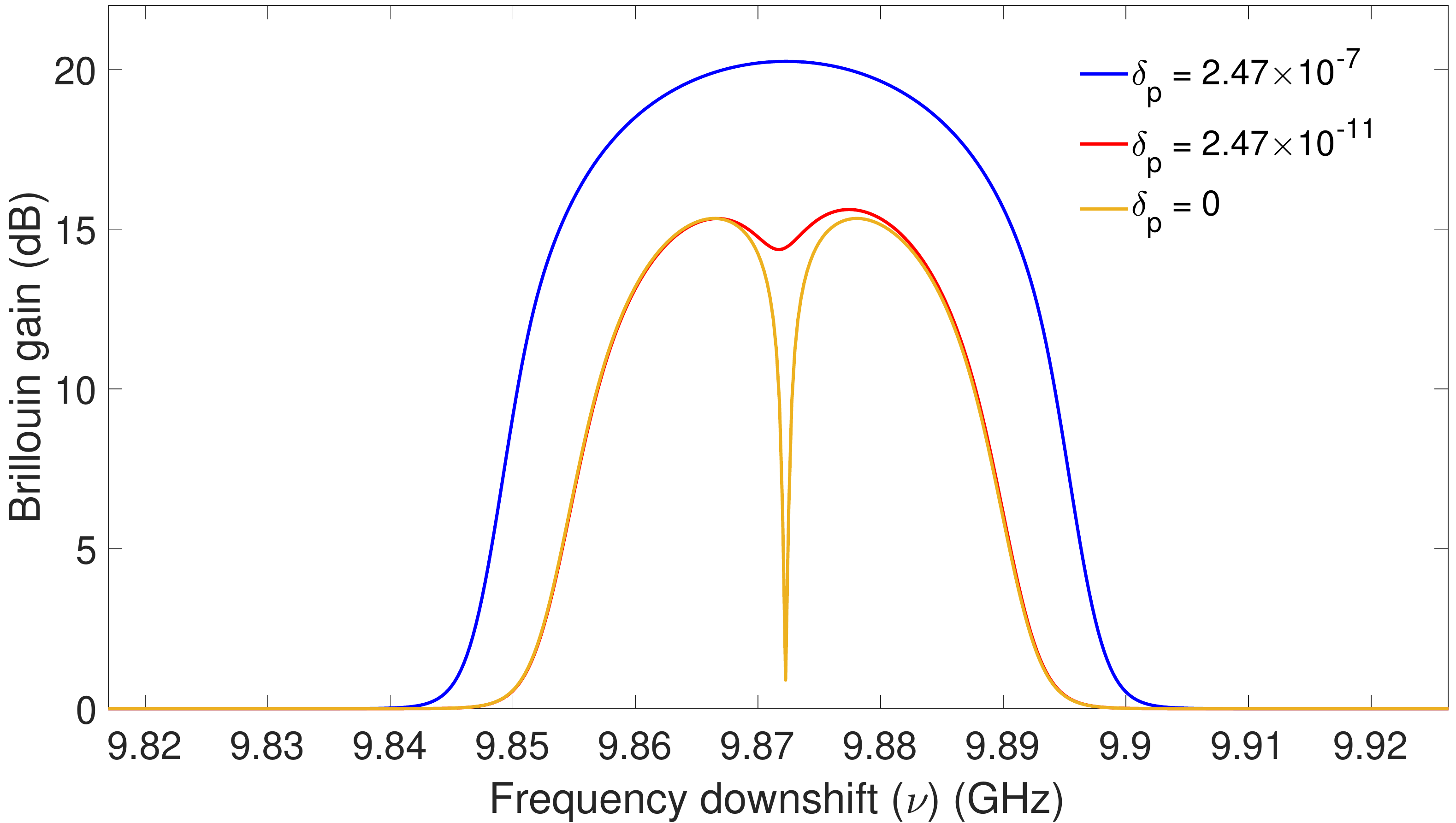}
    \caption{Simulation result showing the effect of change in input polarization on the SBS gain spectrum of SBF. Refer to Eq. \eqref{overlap with pump eigenmode} for the definition of $\delta_p$. For all the cases, the input signal polarization is aligned to $\boldsymbol{S_1}(\nu_p - \nu_B)$, $L$ = 900 m, $L_t$ = 3 mm, $L_b(\nu_p)$ = 26 mm, $\nu_p = 192.97$ THz, $\nu_B$ = 9.8723 GHz, $P_{pump}$ = 14.75 dBm, $P_{sig} = -$12 dBm. Fiber loss is not considered for this simulation.}
    \label{fig:pump polarization misalignment}
\end{figure}

For a fixed input signal polarization $\boldsymbol{S_1}(\nu_s)$, when the pump polarization is deviated from the eigenmode $\boldsymbol{P_2}(\nu_p)$, the signal field experiences a high Brillouin gain (Fig. \ref{fig:pump polarization misalignment}).
\begin{figure}[H]
    \centering
    \includegraphics[scale=0.3]{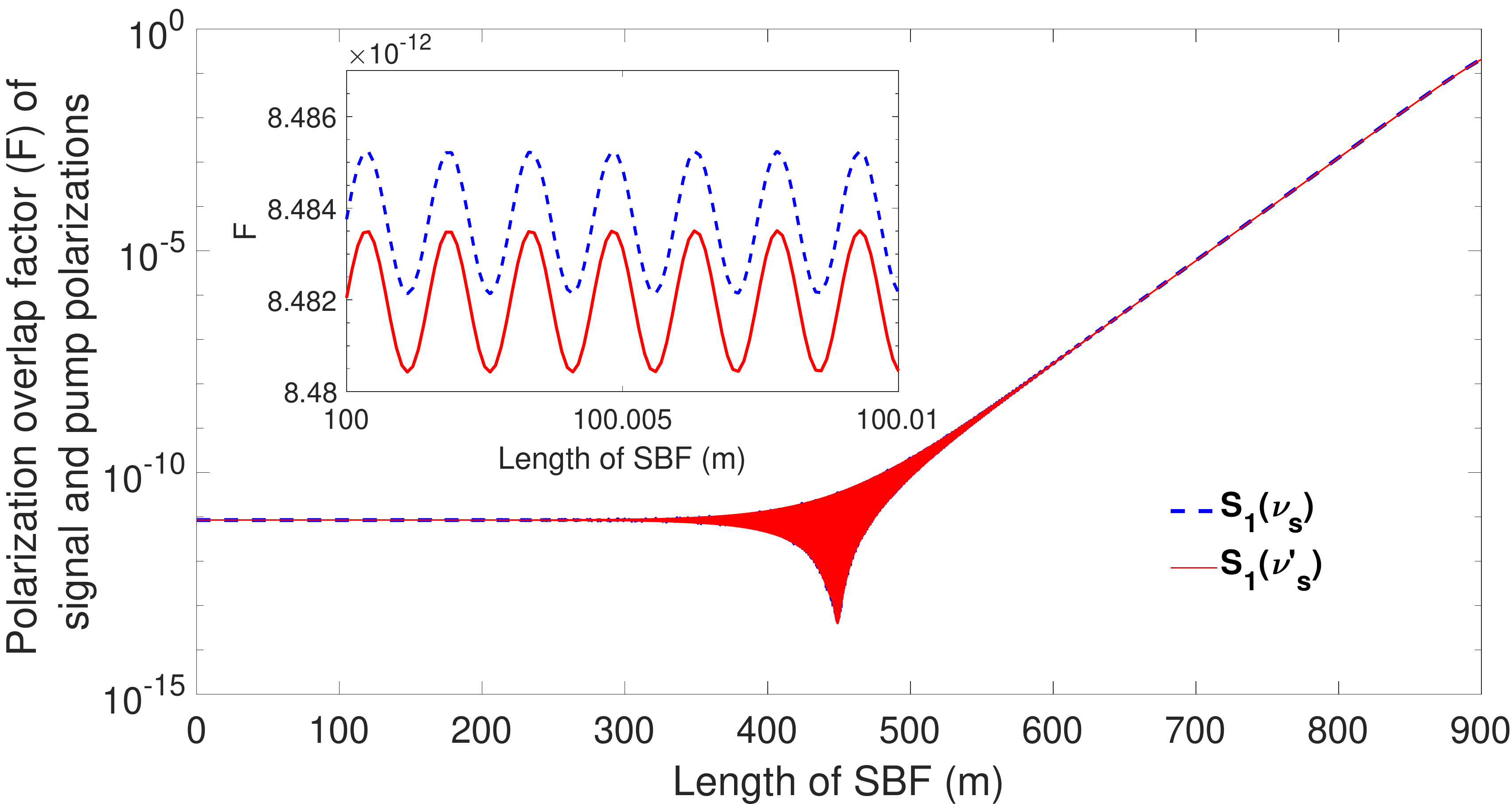}
    \caption{Polarization overlap factor (F) of signal and pump polarization states as a function of length of SBF when eigenmodes $\boldsymbol{S_1}(\nu_s)$ and $\boldsymbol{S_1}(\nu'_s)$ are launched at their respective frequencies $\nu_s$ and $\nu'_s$ ($\nu'_s = \nu_s$ + 1 MHz, and $\nu_s = \nu_p - \nu_B$) for a constant $P_{pump}$ (14.75 dBm). The inset is a zoomed in version near $z$ = 100 m, showing slight polarization oscillation of a non-eigenmode. For both the cases, the input pump polarization is aligned to $\boldsymbol{P_2}(\nu_p)$, $P_{sig} = -$12 dBm, $L_t$ = 3 mm, $L_b(\nu_p)$ = 26 mm, $\nu_p = 192.97$ THz, and $\nu_B$ = 9.8723 GHz. Fiber loss is not considered for this simulation.}
    \label{fig:signal eigenmodes polarization pulling}
\end{figure}
\squeezeup
When pump is launched in eigenmode $\boldsymbol{P_2}(\nu_p)$, and the signal is launched in eigenmode $\boldsymbol{S_1}(\nu_s)$, the signal experiences weak polarization pulling effect. It should be noted that even if signal frequency is changed, say from $\nu_s$ to $\nu'_s$ (for e.g., $\nu'_s = \nu_s$ + 1 MHz), as long as for $\nu'_s$ the signal is launched in its eigenmode $\boldsymbol{S_1}(\nu'_s)$, the effect of weak gain pulling is nearly the same on $\nu_s$ and $\nu'_s$ (see Fig. \ref{fig:signal eigenmodes polarization pulling}).
\section{Role of Polarization-Dependent Gain}
We have stated in the main text that the combination of elliptical birefringence and polarization-dependent gain produces the ultra-narrow spectral feature. But is the gain really essential? What if we replace polarization-dependent gain with polarization-dependent loss? Here, we compare the case of SBF with Brillouin gain with the case of the same SBF with only passive polarizers (serves as polarization-dependent loss elements). 

Note that combining birefringent elements with polarizers essentially produces birefringent filters, which are widely used as spectral filters. In particular, \v{S}olc filters \cite{vsolc1965birefringent,evans1949birefringent} use birefringent elements with their axes rotated at a fix rate along the beam propagation direction, which looks like a discrete analogue to SBF. 
\begin{figure}[H]
    \centering
    \includegraphics[scale=0.31]{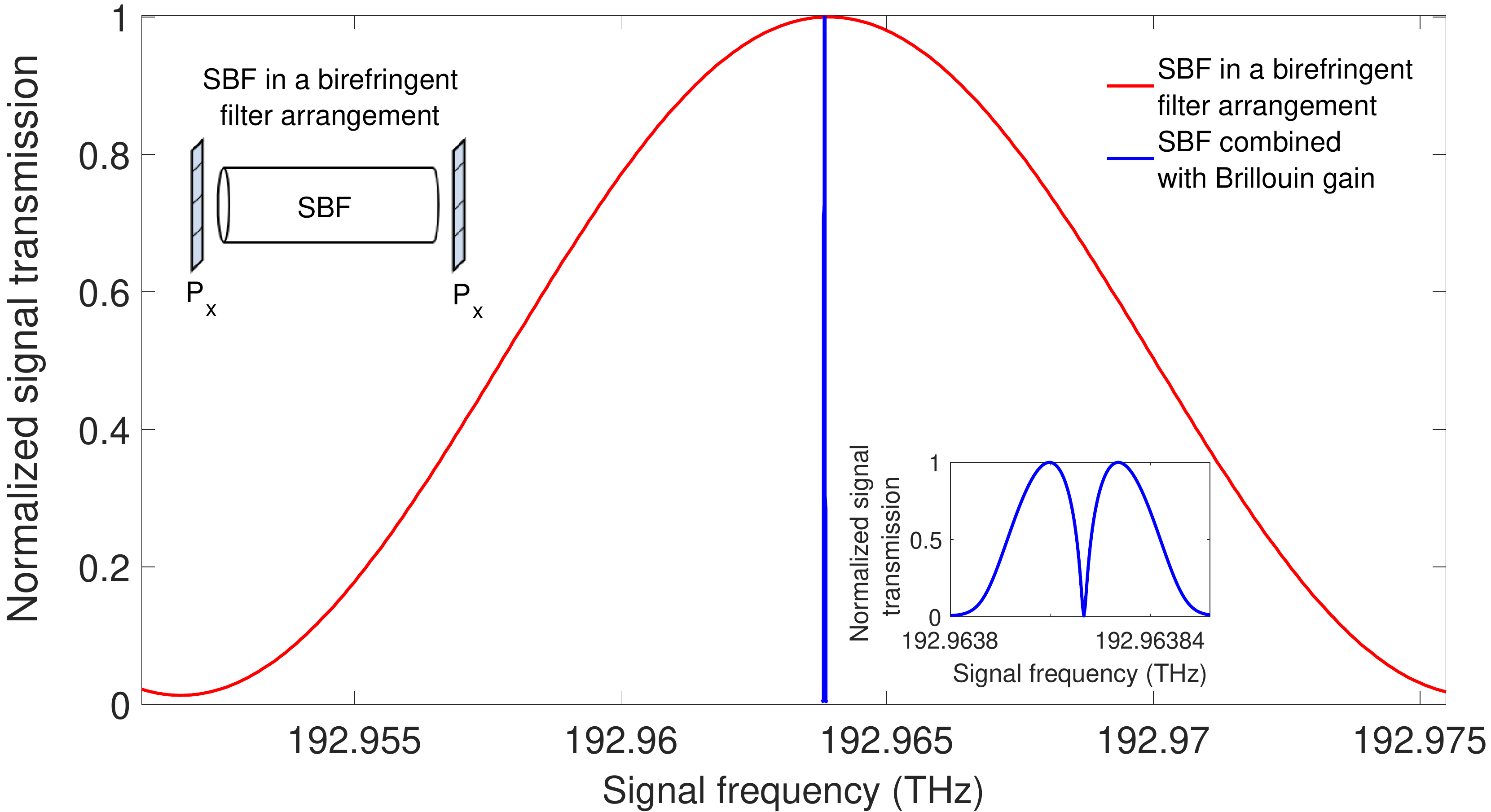}
    \caption{Comparison of the SBS gain spectrum of SBF (blue) with the transmission spectrum of SBF (red) placed in a \v{S}olc filter like arrangement. The configuration of the \v{S}olc filter like arrangement is shown in the upper inset, where $P_x$ are the horizontal polarizers. The lower inset shows a zoomed in version of the SBS gain spectrum (blue curve in the center). The SBF parameters for both the arrangements are as following: $L$ = 900 m, and $L_t$ = 3 mm, $L_b(\nu_p)$ = 26 mm, and $\nu_p = 192.97$ THz. Fiber loss is not considered for both of these simulations. For obtaining the SBS gain spectrum, the input signal and pump input polarizations are $\boldsymbol{S_1}$ and $\boldsymbol{P_2}$, respectively.}
    \label{fig:Solc filter comparison}
\end{figure}
\squeezeup
Let us compare the spectral response of SBS in SBF to that of a \v{S}olc-filter-like passive device made of SBF sandwiched in between two polarizers (see Fig. \ref{fig:Solc filter comparison} and the upper inset for the birefringent filter arrangement). Here, the red curve is the spectral response achieved by keeping SBF in a birefringent filter arrangement. The spectral range of this filter is 24.3 GHz, and the full-width at half maximum is 11.9 GHz. The blue curve is the spectral response achieved when the polarizers are replaced by SBS gain. Thus, by replacing the polarization-dependent loss of the polarizers with the polarization-dependent SBS gain, we can obtain a 5 order-of-magnitude narrowing of the passive filter feature because of the positive feedback from the gain medium.
\section{Comparison of SBS Gain Spectra for Co-Polarized and Orthogonally Polarized Eigenmode Inputs}
\begin{figure}[H]
    \centering
    \includegraphics[scale=0.31]{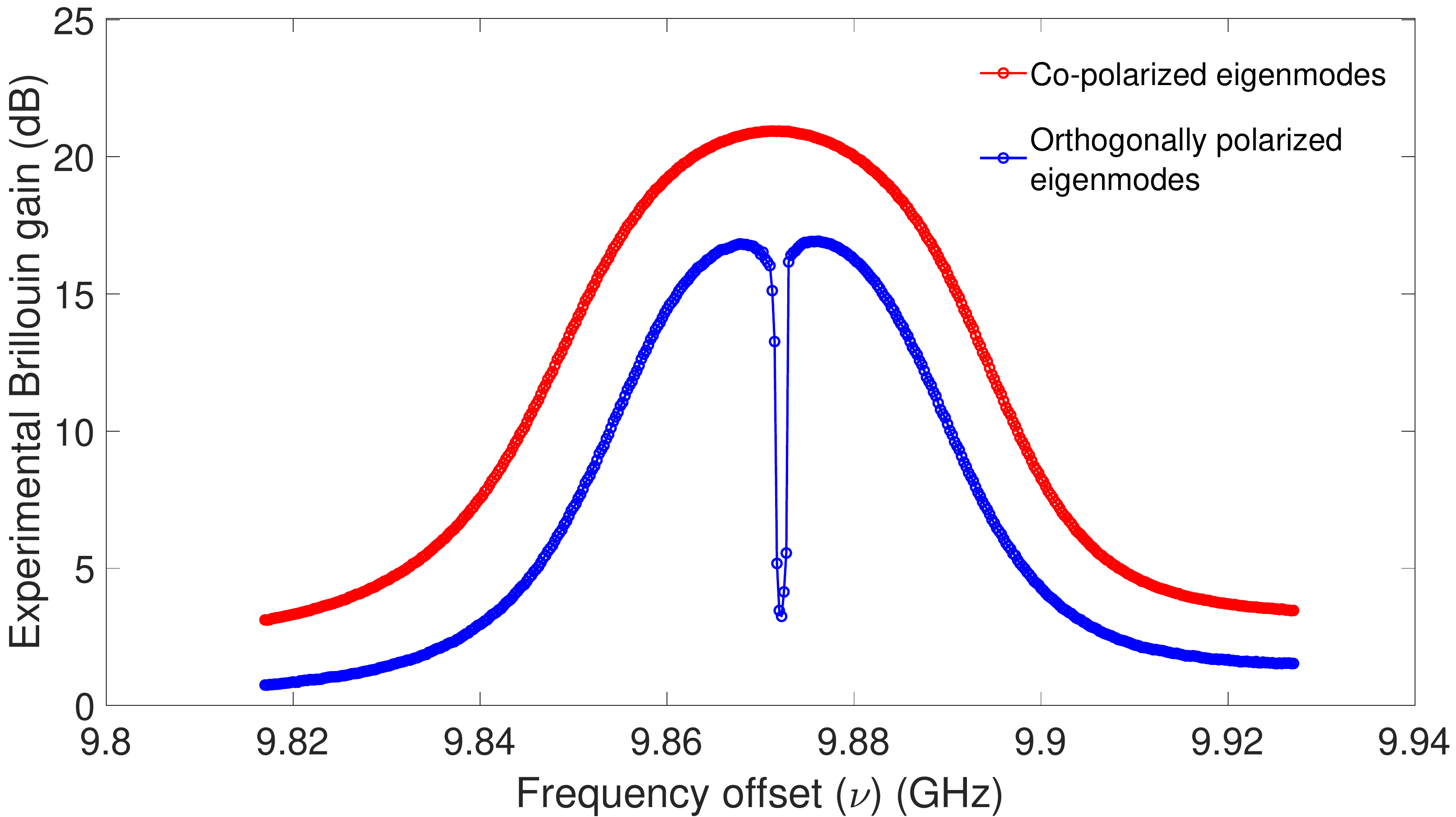}
    \caption{Measured SBS gain spectrum obtained by launching signal and pump as co-polarized (red curve) and orthogonally polarized (blue curve) eigenmodes of SBF. The experimental parameters are: $L$ = 900 m, and $L_t$ = 3 mm, $L_b(\nu_p)$ = 26 mm, $\nu_p = 192.97$ THz, $P_{pump}$ = 16.9 dBm, and $P_{sig} = -$12 dBm.}
    \label{fig:copolarized and orthogonally polarized inputs}
\end{figure}

In the main text, we show that when we launch signal and pump fields as orthogonally polarized eigenmodes ($\boldsymbol{S_1}$ and $\boldsymbol{P_2}$) of SBF, then we observe a spectral dip in the SBS gain spectrum (see Fig. 5b in the main text). In Fig. \ref{fig:copolarized and orthogonally polarized inputs}, we compare the experimental gain spectrum when signal and pump are launched as co-polarized (red curve) and orthogonally polarized eigenmodes of SBF (blue curve). 
\squeezeup
The signal experiences maximum SBS gain when the input signal and pump fields are co-polarized eigenmodes, and contrary to the case of orthogonally polarized case, no spectral dip is observed in the gain spectrum. This observation excludes the possibility of spectral hole burning (SHB) \cite{takushima1995spectral,kovalev2000observation,kovalev2002waveguide,stepien2002origin} being the mechanism behind the dip, because SHB is expected to be more pronounced when gain is higher, which is contrary to our observation  \cite{kovalev2000observation,kovalev2002waveguide}.
\section{Homogeneity of the Observed SBS Gain Spectrum}
Our readers may question the uniformity of our SBF over such a large length ($L$ = 900 m). Here we show that the SBS gain shape obtained experimentally can be closely modelled by a Lorentzian lineshape function.

A homogeneously broadened SBS gain spectrum has a Lorentzian shape, whereas for an inhomogeneously broadened spectrum, the gain shape is expected to resemble more of a Gaussian shape \cite{agrawal2000nonlinear,kovalev2002waveguide}. A higher homogeneity of the SBS gain spectrum suggests that the fiber is more uniform \cite{agrawal2000nonlinear,kovalev2002waveguide}.

\begin{figure}[H]
    \centering
    \includegraphics[scale=0.39]{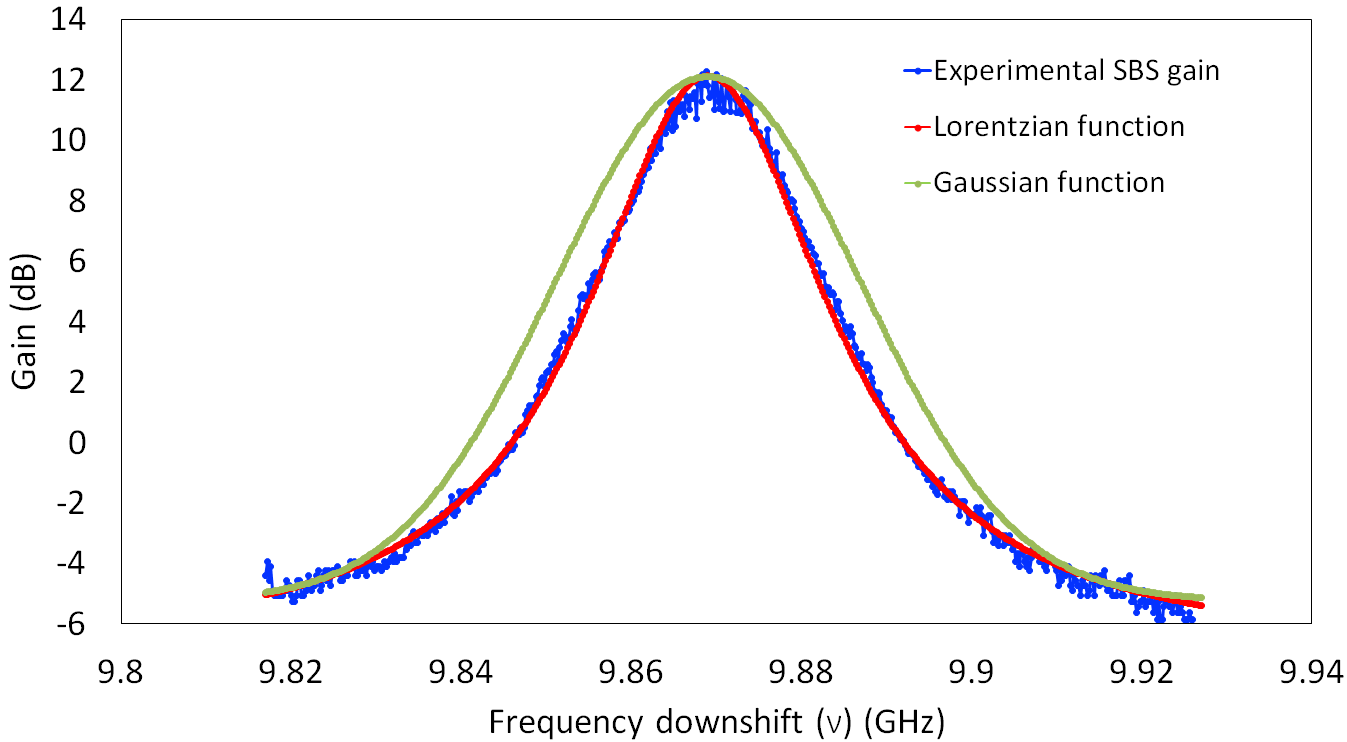}
    \caption{Comparison of the experimental SBS gain spectrum of SBF with Lorentzian and Gaussian gain lineshape functions. The experimental parameters are: $L$ = 900 m, $L_t$ = 3 mm, $L_b(\nu_p)$ = 26 mm, $\nu_p = 192.97$ THz, $P_{pump}$ = 16.9 dBm, and $P_{sig}$ = $-$28.2 dBm. The fitting function has a form of $e^{-\alpha L}e^{kg}$, where $\alpha$ is the loss, $k$ is a real constant, and $g$ is the lineshape function. For Lorentzian fitting function, $g = \frac{\left(\frac{\Gamma}{2}\right)^2}{\left(\Omega-\Omega_0\right)^2+\left(\frac{\Gamma}{2}\right)^2}$, and for Gaussian fitting function, $g = \frac{1}{\frac{\Gamma}{2\sqrt{2ln2}}\sqrt{2\pi}}\ e^{-\frac{1}{2}\left(\frac{\Omega-\Omega_0}{\frac{\Gamma}{2\sqrt{2ln2}}}\right)^2}.$  $\Gamma$ is the bandwidth, $\Omega$ is the center frequency, and $\Omega_0$ is the frequency offset. Optimal values of $\alpha$, $k$, and $\Gamma$ is used to yield the desired fit to the experimental data}
    \label{fig:Gain comparison with Lorentzian}
\end{figure}

In Fig. \ref{fig:Gain comparison with Lorentzian}, we fit the SBS gain spectrum with Lorentzian and Gaussian lineshape functions. We use a low signal power ($P_{sig}$ = $-$28.2 dBm) so that the SBS gain shape is not significantly distorted by pump depletion, gain saturation, or ASE noise in the signal. The input signal and pump polarizations are co-polarized.
The Lorentzian function fits the experimental result well, whereas the Gaussian function does not, indicating that homogeneous gain broadening is the dominant broadening mechanism. This suggests that the fiber is nearly uniform in nature.
\section{Variation of Dip Depth for an Optimal BTR of 1}
\begin{figure}[H]
    \centering
    \includegraphics[scale=0.4]{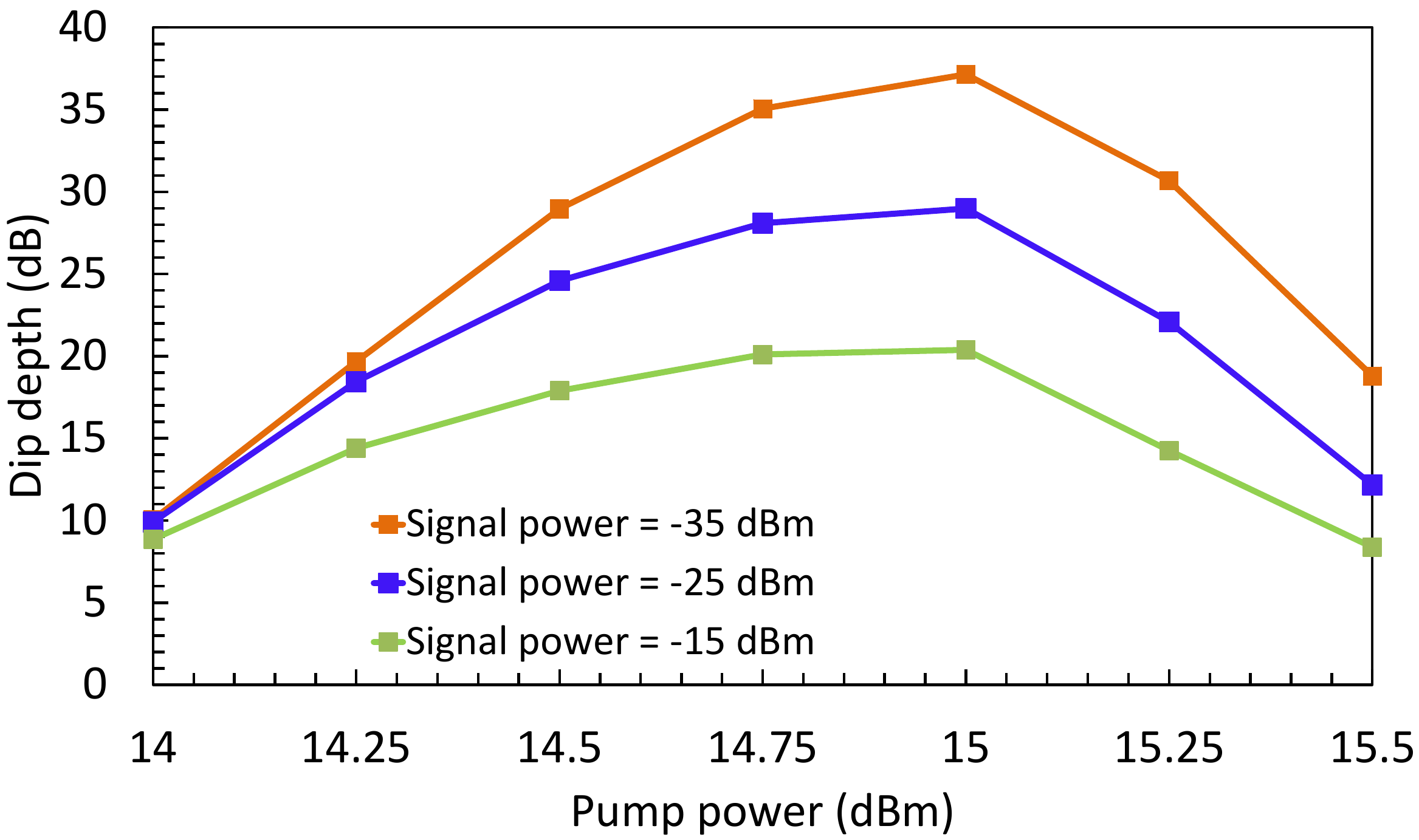}
    \caption{Simulation result showing a variation of the dip depth with a change in $P_{sig}$ and $P_{pump}$ for an optimal BTR of 1. The simulation parameters are: $L$ = 900 m, $L_b(\nu_p)$ = 0.26 mm, $\nu_p = 192.97$ THz, and $L_t$ = 0.26 mm. Fiber loss is not considered for these simulations. The symbols are simulation data points, and the lines are guides to the eye}
    \label{fig:dip depth variation}
\end{figure}

Figure 3b in the main article shows that dip depth is maximized at BTR = 1, for given signal and pump powers. Here, we show that, given BTR = 1, dip depth can be further increased by an optimal combination of $P_{sig}$ and $P_{pump}$. As shown in Fig. \ref{fig:dip depth variation}, the dip depth increases with a decrease in $P_{sig}$. This is because a lower $P_{sig}$ results in weaker gain saturation, and therefore higher gain away from the dip. The dip depth increases with an increase in $P_{pump}$, reaches a maximum of 37 dB for a $P_{pump}$ = 15 dBm and $P_{sig}$ = $-$35 dBm, and then decreases with a further increase in $P_{pump}$. For high input pump powers (>15 dBm), the signal experiences higher polarization pulling of the pump, even at the dip frequency (at the dip frequency both signal and pump are nearly orthogonal since they are eigenmodes), and this leads to a decrease in dip depth. For low input pump powers ($\leq$ 14 dBm), the SBS gain in signal is less than 10 dB, and thus the pump depletion and gain saturation is is negligible over the range of the signal powers ($-$35 dBm to $-$15 dBm). Therefore, for these pump powers, the dip depth is nearly same for this range of input signal powers. 
\squeezeup
\bibliography{Supplementary_material}